\documentclass[12pt]{iopart}

\begin{document}

\title[Stability analysis of the cosmological solutions with induced gravity and scalar field on the brane]
{Stability analysis of the cosmological solutions with induced
gravity and scalar field on the brane}

\author{{ Kourosh Nozari \footnote{knozari@umz.ac.ir}}\quad,{ F. Rajabi
 \footnote{f.rajabi@stu.umz.ac.ir}}\quad and\quad  { K. Asadi  \footnote{k.asadi@stu.umz.ac.ir}}}

\address{Department of Physics, Faculty of Basic Sciences,\\ University of Mazandaran, P. O. Box 47416-95447,\\ Babolsar, IRAN}

\begin{abstract}
We study cosmological dynamics and phase space of a scalar field
localized on the DGP brane. We consider both the minimally and
nonminimally coupled scalar quintessence and phantom fields on the
brane. In the nonminimal case, the scalar field couples with induced
gravity on the brane. We present a detailed analysis of the critical
points, their stability and late-time cosmological viability of the
solutions in the phase space of the model.
\end{abstract}

\maketitle

\section{Introduction}
In the revolutionary braneworld viewpoint, our universe is a
$3$-brane embedded in an extra dimensional bulk. Standard matter and
all interactions are confined on the brane; only graviton and
possibly non-standard matter are free to probe the full bulk
\cite{1}. Based on the braneworld viewpoint, our universe may
contain many more dimensions than those we experience with our
senses. The most compelling reasons to believe in extra dimensions
are that they permit new connections between physical properties of
the observed universe and suggest the possibility for explaining
some of its more mysterious features. Extra dimensions can have
novel implications for the world we see, and they can explain
phenomena that seem to be mysterious when viewed from the
perspective of a three-dimensional observer. Even if one is doubtful
about string theory due to, for instance, its huge number of
landscapes, recent researches have provided perhaps the most
compelling argument in the favor of extra dimensions: a universe
with extra dimensions might contain clues to physics puzzles that
have no convincing solutions without them. This reason alone makes
extra dimensional theories worthy of investigation. In this
streamline, the braneworld models that are inspired by ideas from
string theory provide a rich and interesting phenomenology, where
higher-dimensional gravity effects in the early and late universe
can be explored, and predictions can be made in comparison with
high-precision cosmological data.  Even for the simplest models of
Randall-Sundrum (RS) \cite{2} and Dvali-Gabadadze-Porrati (DGP)
\cite{3}, braneworld cosmology brings new implications on the
inflation and structure formation. Also it brings new ideas for dark
energy and opens up exciting prospects for subjecting M-theory ideas
to the increasingly stringent tests provided by high-precision
astronomical observations. At the same time, braneworld models
provide a rich playground for probing the geometry and dynamics of
the gravitational field and its interaction with matter. In these
respects, the DGP braneworld model is a scenario that gravity is
altered at immense distances by the excruciatingly slow leakage of
gravity off our $3$-brane universe. In this braneworld scenario, the
bulk is considered as empty except for a cosmological constant and
the matter fields on the brane are considered as responsible for the
evolution on the brane \cite{3}. The self-accelerating DGP branch
explains late-time speed-up by itself without recourse to dark
energy or other mysterious components \cite{4}. Even the normal DGP
branch has the potential to realize an effective phantom phase via
dynamical screening of the brane cosmological constant \cite{5}.

Scalar fields play a crucial role in modern cosmology, both in
models of the early universe and late-time acceleration. Scalar
fields provide also a simple dynamical model for matter fields in a
braneworld and dark energy models. In the early universe, inflaton
as a scalar field provides the required basis of the some
well-established inflation models. Also at late time, dark energy
models based on dynamical scalar fields have been studied
extensively in recent years \cite{6}. In braneworld models, the
existence of a scalar field on the brane provides a variety of
possibilities that brings the corresponding theory to explain some
novel properties. In fact, a particular form of the bulk or brane
matter is a scalar field. In the context of braneworld induced
gravity, it is natural to consider a non-minimal coupling of the
scalar field and induced Ricci curvature on the brane. The resulting
theory can be thought of as a generalization of the Brans–Dicke type
scalar-tensor gravity in a braneworld context \cite{7}. As has been
pointed in \cite{8}, the introduction of the non-minimal coupling
(NMC) is not just a matter of taste: the NMC is forced upon us in
many situations of physical and cosmological interest. For instance,
NMC arises at the quantum level when quantum corrections to the
scalar field theory are considered. Even if for the classical,
unperturbed theory this NMC vanishes, it is necessary for the
renormalizability of the scalar field theory in curved space. In
most theories used to describe inflationary scenarios, it turns out
that a non-vanishing value of the coupling constant is inevitable.
In general relativity, and in all other metric theories of gravity
in which the scalar field is not part of the gravitational sector,
the coupling constant necessarily assumes the value of $\frac{1}{6}$
\cite{8}. Therefore, it is natural to incorporate an explicit NMC
between the scalar field and the Ricci scalar in the inflationary
paradigm and also in scalar fields models of dark energy. In
particular the effect of this NMC in a DGP-inspired braneworld
cosmology has been studied by some authors (see \cite{7} and also
\cite{9}).

There are several studies focusing on braneworld models with
brane/bulk scalar fields. Some of these studies concentrate on the
bulk scalar fields minimally or nonminimally coupled to the bulk
Ricci scalar \cite{10}. The scalar field minimally or non-minimally
coupled to gravity on the brane are studied by some authors
\cite{11,12,13,14}. In \cite{14}, the authors are studied the
self-accelerating solutions in a DGP brane with a scalar field
trapped on it within a dynamical system perspective. They have shown
that the dynamical screening of the scalar field self-interaction
potential occurring within the Minkowski cosmological phase of the
DGP model mimics 4D phantom behavior and is an \emph{attractor}
solution for a constant self-interaction potential. But, this is not
the case necessarily for an exponential potential. For exponential
potential, they have shown that gravitational screening is not even
a critical point of the corresponding autonomous system of ordinary
differential equations. Along with this pioneer work, we consider a
scalar field trapped on the DGP brane and we suppose this scalar
field is non-minimally coupled to induced gravity on the brane. We
study cosmological dynamics on the normal branch of the scenario
within a phase space approach with both quintessence and phantom
fields on the brane. We provide a phase space analysis of each model
through a detailed study of the fixed points, their stability and
cosmological viability of the solutions. We also study the classical
stability of the solutions in each case in the
$w_{\varphi}-w_{\varphi}'$ phase-plane. Our study, in comparison
with existing literature in this field, provides a complete
framework and contains several aspects of the problem not been
considered yet. Since the self-accelerating DGP branch has ghost
instabilities, our study here is restricted just to the normal DGP
branch of the models. While the normal branch of a pure DGP setup
has not the potential to explain the late-time cosmic speed-up and
crossing of the phantom divide, we show that with a scalar field on
the brane there are several new possibilities in the favor of these
observationally supported issues.

We note that our motivation to study this model is as follows: as we
have indicated above, observations support (at least mildly) that
the equation of state parameter of dark energy has crossed the
cosmological constant line ($w=-1$) in recent past (at redshift
$z\sim0.25$). It is impossible to realize this feature with a
quintessence or phantom field minimally coupled to gravity in
standard 4-dimensional theory \cite{6}. Although the original DGP
model was proposed to realize accelerated expansion of the universe
in a braneworld setup, the self-accelerating branch of the DGP
cosmological solutions has ghost instability. It is impossible also
to cross the phantom divide line without a scalar field in the
self-accelerating DGP branch \cite{11}. The normal DGP solution has
no ghost instability, but it cannot explain accelerated expansion
and crossing of the phantom divide. It has been shown that
localizing a scalar field on the normal DGP setup realizes these
features \cite{11}. It is possible also to incorporate extra degrees
of freedom on the braneworld setup to have more successful models
(see for instance \cite{7}. These extra degrees provide new
facilities and richer cosmological history on the brane, a part of
which is related to the wider parameter space. On the other hand,
considering just a cosmological constant on the brane, although
explains accelerated expansion through dynamical screening of the
brane cosmological constant, it has not the potential to explain
crossing of the phantom divide \cite{5}. In this paper we have shown
that a scalar field, minimally or nonminimally coupled to induced
gravity on the brane has the potential to fill these gaps. We stress
that all of the accelerated phases obtained in this study belong to
the normal DGP branch of the model. Our study, based on the phase
space analysis, provides the most complete treatment of the issue in
the field. We have provided a complete analysis of the problem
focusing on all possible details.

\section{Cosmological dynamics of a minimally coupled scalar field on the DGP brane}

DGP braneworld scenario has attracted a lot of attention through
these years. Although this scenario has very interesting
phenomenological aspects, it suffers from shortcomings such as ghost
instabilities in its self-accelerating branch of the solutions.
Fortunately the normal, non-self-accelerating branch of the DGP
cosmological solutions has no ghost instabilities. For this reason
we consider a scalar field on the DGP brane and we analyze
cosmological dynamics of the \emph{normal DGP} solutions in this
setup. To begin and in order to explain the frame of our analysis,
we start with the case of a minimally coupled quintessence field on
the normal DGP branch. We note that this problem has been considered
previously in \cite{11,12,14}. Especially, in \cite{14} the authors
have presented a detailed study of the problem with just a minimally
coupled scalar field on the brane. We firstly study quintessence
field with more details and more enlightening analysis than
\cite{14} both in calculations and corresponding analysis of the
phase space points. Then we extend our study to the minimally
coupled phantom fields, nonminimally coupled quintessence fields and
finally non-minimally coupled phantom fields on the brane. In each
step, we provide a detailed analysis of the model in phase space and
within a dynamical system approach. We study the late-time
cosmological viability of the solutions and their stability in each
step. We also investigate the classical stability of the solutions
in $w_{\varphi}-w_{\varphi}'$ phase-plane.

\subsection{Minimally coupled quintessence field on the brane}

A minimally coupled quintessence field on the normal DGP within a
phase space approach first has been considered in \cite{11}. There,
the authors have considered the cosmological evolution of a QDGP
model in a phantom-like prescription. They have defined an effective
cosmological constant that increases by time evolution of the Hubble
parameter and therefore realizes a phantom-like behavior through
dynamical screening of the brane cosmological constant. Here we
adopt another strategy: we focus on the phase space coordinates
instead of the brane cosmological constant similar to strategy
adopted in \cite{14}. Although in this step the results are the
same, but our analysis in this subsection introduces the notation,
conventions and general framework of our procedure. In addition,
some novel ingredients such as $w_{\varphi}-w_{\varphi}'$
phase-plane stability analysis, more detailed calculations and more
enlightening plots are presented.

The action of an induced gravity braneworld model can be written
as follows (we use the sign convention of \cite{6})
\begin{equation}
\fl{\cal{S}}={\frac{-M^3_{5}}{2}}\int d^5
X\sqrt{-g}R_{5}-{\frac{\mu^2}{2}}\int d^4 x \sqrt{-h}R_4 +\int d^4
x\sqrt{-h}{\cal{L}}_{m}+{\cal{S}}_{GH}\,,\label{eq1}
\end{equation}
where $g_{ab}$ is the metric of the bulk manifold and $h_{\mu\nu}$
is the induced metric on the brane. $R_5$ and $ R_4$ denote the
$5$ and $4$ dimensional Ricci scalars respectively, and
${\cal{L}}_{m}$ is the matter Lagrangian confined on the brane.
${\cal{S}}_{GH}$ is the Gibbons-Hawking boundary action which is
required in order to apply the boundary conditions properly. In
this induced gravity braneworld setup, the ratio of the two
scales, the $4$-dimensional Planck mass $\mu$ and its
$5$-dimensional counterpart $M_{5}$, defines the DGP crossover
scale as\[
r_c=\frac{\mu^2}{2M^3_{5}} \,,
\]
which determines the behavior of gravity in different distance
scales on the brane. Adopting a FRW line element, the cosmology of
the model is based on the following Friedmann equation \cite{4,11}
\begin{equation}
H^2+{\frac{K}{a^2}}=\Bigg(\sqrt{\frac{\rho}{3\mu^2}+{\frac{1}{4r^2_c}}}+{\frac{1}{2r_c}}\Bigg)^2\,,\label{eq2}
\end{equation}
where $\rho$ is the energy density of the total cosmic fluid on
the brane and consists of the energy densities of the scalar field
and ordinary matter on the brane. We consider the flat geometry in
which $(K=0)$, so the above equation reduces to
\begin{equation}
H^2+{\frac{H}{r_c}}={\frac{\rho}{3\mu^2}}\,,\label{eq3}
\end{equation}
for the \emph{normal} DGP branch of the scenario. When the Hubble
length $H^{-1}$ is much smaller than $r_{c}$, which stands for the
early time, the term $\frac{H}{r_{c}}$ can be ignored relative to
the first term on the left hand side of \eref{eq3}. This term
becomes important on scales comparable to the crossover scale when
$H^{-1}$ is larger than $r_{c}$, which corresponds to the late-time
of the universe evolution. We emphasize that in which follows we
focus just on the normal, non-self-accelerating DGP branch of the
solutions. This is because the normal branch has no ghost
instabilities. Nevertheless, the pure normal branch cannot explain
the late-time acceleration without additional components on the
brane. However, in the presence of a scalar field on the brane, it
is possible to realize the late-time cosmic speed-up even in the
normal DGP branch \cite{11,14}. This is the reason why we considered
an extension of the DGP setup with a scalar field on the brane.

By differentiation of \eref{eq3} with respect to the cosmic time,
we obtain an equation for evolution of the Hubble parameter
\begin{equation}
\dot {H}={\frac{-(\rho+p)}{2\mu^2}}\Big(1+{\frac{1}{2r_{c}
H}}\Big)^{-1}\,,\label{eq4}
\end{equation}
where we have used the continuity equation on the brane as
$\dot{\rho}+3H(\rho+p)=0$. We note that the negativity of
$\dot{H}$ ensures the phantom-like behavior on the brane
\cite{11}. The dynamics of the scalar field localized on the brane
is given by the following Klein-Gordon equation
\begin{equation}
\ddot{\varphi}+3H{\dot{\varphi}}=-\frac{dV}{d\varphi}\,.\label{eq5}
\label{eq5}
\end{equation}
The energy density and pressure of the total matter localized on
the brane are given by
\begin{equation}
\rho=\rho_{\varphi}+\rho_m=\frac{1}{2} {\dot{\varphi}}^2
+V(\varphi)+\rho_m\,, \label{eq6}
\end{equation}
\begin{equation}
p=p_{\varphi}+p_m=\frac{1}{2} {\dot{\varphi}}^2 -V(\varphi)+w_m
\rho_m  \label{eq7}
\end{equation}
respectively. To translate our equations of the cosmological
dynamics in the language of the autonomous dynamical system, we
introduce the following dimensionless quantities

\begin{equation}
\eqalign{x_1={\frac{\dot{\varphi}}{\sqrt{6}\mu H}}\,,\qquad
x_2={\frac{\sqrt{V}}{\sqrt{3}\mu H}}\,, \qquad
x_3={\frac{\sqrt{\rho_m}}{\sqrt{3}\mu H}}\,,\\
x_4={\frac{1}{\sqrt{2r_c H}}}\,,\qquad \lambda=-{\frac{V^\prime
\mu}{V}}\,, \qquad\quad \Gamma={\frac{V
V^{\prime\prime}}{{V^{\prime}}^2}}\,,}\label{eq8}
\end{equation}
where a prime marks differentiation with respect to the scalar
field, $\prime\equiv\frac{d}{d\varphi}$. Through this paper we
consider the exponential potential of the scalar field as
$V(\varphi)=V_{0}e^{{-\lambda \varphi}/{\mu}}$. This potential
corresponds to a constant $\lambda$ and gives $\Gamma=1$. The case
of a constant potential (as has been studied separately in
\cite{14}), is a special case of this potential. We note that the
above dimensionless quantities have explicit physical origin:
$x_{1}^{2}$ is related to the kinetic energy of the field,
$x_{2}^{2}$ is related to the potential energy of the scalar field,
$x_{3}^{2}$ is related to the ordinary matter density on the brane,
$x_{4}^{2}$ reflects the DGP character of this setup, and $\lambda$
and $\Gamma$ are actually the slow-roll parameters of the model. Our
main equations are the Friedmann equation that appears as a
constraint, the Klein-Gordon equation and the continuity equation on
the brane. These equations with the above dimensionless quantities
provide the basis of our dynamical system analysis.

By using the above dimensionless quantities, we rewrite \eref{eq5},
\eref{eq6} and \eref{eq7} in the following form

\begin{equation}
\ddot{\varphi}=-3 \mu H^2(\sqrt{6}x_1-\lambda x_2^2)\,,\label{eq9}
\end{equation}

\begin{equation}
\rho=3\mu^2 H^2 (x_1^2+x_2^2+x_3^2)\,,\label{eq10}
\end{equation}

\begin{equation}
p=3\mu^2 H^2 (x_1^2-x_2^2+w _m x_3^2)\,.\label{eq11}
\end{equation}

The effective equation of state parameter in this case is given by
\begin{equation}
w_{eff}=\frac{x_1^{\,\,2}-x_2^{\,\,2}+w_{m}x_3^{\,\,2}}{x_1^{\,\,2}+x_{2}^{\,\,2}+x_{3}^{\,\,2}}\,.\label{eq12}
\end{equation}

We can obtain the exact cosmological solutions at the critical
points by using the following relation
\begin{equation}
\dot{H}=-\frac{3}{2}
\frac{[2x_1^2+(1+w_m)x_3^2]}{1+x_4^2}H^2\,.\label{eq13}
\end{equation}
In each fixed point, we can rewrite this relation in the following
form
\begin{equation}
\dot {H}=-\frac{1}{\alpha}H^{2}\,,\label{eq14}
\end{equation}
where by definition
\begin{equation}
\alpha=\frac{2(1+x_{4}^{\,\,2})}{3\Big(2x_1^{\,\,2}+(1+w_{m})x_3^{\,\,2}\Big)},\,\,\quad
\alpha\neq 0 \,.\label{eq15}
\end{equation}
An integration of \eref{eq14} with respect to the cosmic time gives
\begin{equation}
a(t)=a_{0}(t-t_{0})^{\alpha}\,\ \label{eq16}
\end{equation}
which corresponds to an accelerating phase if $\alpha >1$. Note that
equation (16) is valid only in a small neighborhood around the
critical point where $\alpha$ can be considered to be nearly
constant. Now the Friedmann constraint equation in terms of the
dimensionless quantities becomes
\begin{equation}
x_1^2+x_2^2+x_3^2-2x_4^2=1\,.\label{eq17}
\end{equation}
So, the allowable region of the phase space is actually outside of
a unit sphere defined as $x_{1}^2+x_{2}^2+x_{3}^2=1$. Now the
autonomous dynamical equations are given as follows
\begin{equation}
\frac{d{x}_{1}}{dN}=-3x_1+{\frac{\sqrt{6}}{2}}\lambda x_2^2+3x_1
\Bigg({\frac{2x_1^{\,\,2}+(1+w_{m})x_{3}^{\,\,2}}{1+x_{1}^{\,\,2}+x_{2}^{\,\,2}+x_{3}^{\,\,2}}}\Bigg)\,,\label{eq18}
\end{equation}

\begin{equation}
\frac{d{x}_{2}}{dN}=-\frac{\sqrt{6}}{2}\lambda x_1
x_2+3x_2\Bigg({\frac{2x_1^{\,\,2}+(1+w_{m})x_{3}^{\,\,2}}{1+x_{1}^{\,\,2}+x_{2}^{\,\,2}+x_{3}^{\,\,2}}}\Bigg)\,,\label{eq19}
\end{equation}

\begin{equation}
\frac{d{x}_{3}}{dN}=-\frac{3}{2}(1+w_{m})x_{3}+3x_3\Bigg({\frac{2x_1^{\,\,2}+(1+w_{m})x_{3}^{\,\,2}}{1+x_{1}^{\,\,2}+x_{2}^{\,\,2}+x_{3}^{\,\,2}}}\Bigg)\,,\label{eq20}
\end{equation}
where by definition, $N=\ln a(t)$. The eigenvalues of the Jacobian
matrix are as follows\\

$\bullet$ points 1a , 1b :$$\alpha_{1}=\frac{3}{2}\gamma,\quad
\alpha_{2}=\frac{3}{2}\gamma,\quad
\alpha_{3}=-\frac{3}{2}(2-\gamma)\,.$$

$\bullet$ point 2a :$$\alpha_{1}=3,\quad
\alpha_{2}=\frac{-\sqrt{6}}{2}\lambda+3,\quad
\alpha_{3}=\frac{3}{2}(2-\gamma)\,.$$

$\bullet$ point 2b :$$\alpha_{1}=3,\quad
\alpha_{2}=\frac{\sqrt{6}}{2}\lambda+3,\quad
\alpha_{3}=\frac{3}{2}(2-\gamma)\,.$$

$\bullet$ points 3a , 3b :$$\alpha_{1}=\frac{\lambda^{2}}{2},\quad
\alpha_{2}=\frac{\lambda^{2}}{2}-3,\quad
\alpha_{3}=\frac{1}{2}(\lambda^{2}-3\gamma)\,.$$

$\bullet$ points 4a , 4b , 4c , 4d :
$$\alpha_{1}=\frac{3}{2}\gamma,\quad \alpha_{2,3}=\frac{-3}{4}(2-\gamma)\Bigg(1\pm{\sqrt{1-\frac{8\gamma(\lambda^2-3\gamma)}{\lambda^2(2-\gamma)}}}\Bigg)\,,$$
where $\gamma \equiv 1+w_{m}$ is the barotropic index which depends
on the type of ordinary matter on the brane. In what follows, we
consider $0<\gamma<2$.

\Tref{t1} summarizes the complete information about existence,
stability and cosmological characteristics of these phase points. In
this table, $\Omega_{\varphi}=\frac{\rho_{\varphi}}{\rho_{c}}$ where
$\rho_{c}=3\mu^{2}H^{2}$ and $\gamma_{\varphi}=1+w_{\varphi}$. We
note that for point $1$ with $\gamma=0$, there is a stable center.
All other mentioned points have center too, depending on the values
of $\gamma$ and $\lambda$. But these centers are not necessarily
stable. For instance, for point \{(3a),(3b)\}, the center is
unstable.

\begin{table*}
\begin{tiny}
\caption{\label{t1}\small{Location and dynamical character of the
fixed points.}}\label{t1}
\begin{tabular}{|c|c|c|c|c|c|c|c|c|c|} \hline\mr\ns

name & $x_{1c}$ & $x_{2c}$ & $x_{3c}$ & Existence & stability &
$\Omega_{\varphi}$ &$ \gamma_{\varphi}$&$w_{eff}$&$a(t)$\\
\hline\mr\ns

(1a),(1b)& 0 & 0 & $\pm{1}$& $\forall\, \lambda,\, \gamma$  &
saddle point & 0 & undefined &
$\gamma-1$&$a_{0}(t-t_{0})^{2/3\gamma}$\\\hline

(2a)& 1& 0& 0 &$\forall \, \lambda,\, \gamma$ & saddle point for
\quad
$\lambda>{\sqrt{6}}$ & 1 & 2 & 1 &$a_{0}(t-t_{0})^{1/3}$ \\
  &  &  &  & &   unstable node for \quad $\lambda<{\sqrt{6}}$ & & & & \\ \hline

(2b)& -1&0 & 0 &$\forall \, \lambda,\, \gamma$ &  unstable node for \quad$\lambda>{-\sqrt{6}}$ & 1 & 2 & 1 & $a_{0}(t-t_{0})^{1/3}$ \\
 &  &  &  &  & saddle point for \quad$\lambda<{-\sqrt{6}}$ & & & & \\\hline

(3a),(3b)&$\frac{\lambda}{\sqrt{6}}$ &$
\pm(1-\frac{\lambda^2}{6})^{1/2}$ & 0 &$\lambda^2\leq6$ &saddle
point &1&$\frac{\lambda^2}{3}$& $\frac{\lambda^2}{3}-1$&
$a_{0}(t-t_{0})^{2/\lambda^2}$\\ \hline

(4a),(4b) &  &  &  &  & stable node for \quad $3\gamma<\lambda^2<\frac{24\gamma^2}{9\gamma-2}$ & & &&  \\
(4c),(4d)&
${\sqrt{\frac{3}{2}}}{\frac{\gamma}{\lambda}}$&$\pm({\frac{3\gamma(2-\gamma)}{2\lambda^2}})^{1/2}$
& $\pm (1-{\frac{3\gamma}{\lambda^2}})^{1/2}$& $\lambda^2>3\gamma$
& stable spiral for
\quad$\,\,\lambda^2>\frac{24\gamma^2}{9\gamma-2}$ &$
{\frac{3\gamma}{\lambda^2}}$&$\gamma $ & $\gamma-1$
&$a_{0}(t-t_{0})^{2/3\gamma}$\\ \hline\mr\ns
\end{tabular}
\end{tiny}
\end{table*}

For points $\{(1a),(1b)\}$, there is no contribution of the scalar
field and the universe is dominated by other matter fields.
According to the eigenvalues and since $0<\gamma<2$, these points
behave like saddle points in the phase space.

Points $\{(2a),(2b)\}$ are solutions dominated by the kinetic
energy of the scalar field. The contribution of scalar field
potential and the energy densities of other matter fields are
irrelevant for these phases. For these points, we obtain
$\gamma_{\varphi}=2$, which is referred to as a \emph{stiff
matter} equation of state. These points have no late time
acceleration and the stability of them depends on the values of
$\lambda$, so that for $\lambda<\sqrt{6}$ (for point $(2a)$) and
$\lambda>-\sqrt{6}$ (for point $(2b)$), we obtain unstable nodes.
Otherwise they are saddle points.

For critical points $\{(3a),(3b)\}$, the energy density of universe
is dominated by the scalar field's kinetic and potential energies.
For $\lambda^{2}<2$\, we have accelerated phase of expansion, but
this phase is not stable. In these cases, for $\lambda=0$, the
universe is dominated by a cosmological constant.

 The last line of \tref{t1}\, contains four critical points
$\{(4a),(4b),(4c),(4d)\}$, so that depending on the values of
$\gamma$ and $\lambda$, we have stable spirals or stable nodes. Here
$\gamma_{\varphi}$ is equal to the barotropic index of matter,
$\gamma$. This is a reflection of the fact that the exponential
potential used in this framework can give rise to an accelerated
expansion and possesses cosmological scaling solutions in which the
field energy density $\rho_{\varphi}$ is proportional to the fluid
energy density, $\rho_{m}$.

\begin{table*}
\begin{center}
\caption{\label{t2}\small{Location and critical point for
$\gamma_{\varphi}=0$ (that is, $w_{\varphi}=-1$).}} \label{t2}
\begin{tabular}{|c|c|c|c|c|c|c|}\hline\mr\ns

name  & $x_{2c}$ & $x_{3c}$  & $ x_{4c}$ & stability &$w_{eff}$&
a(t) \\ \hline\mr\ns

(1a),(1b) & $0$ & $\pm 1$ & $0$ & unstable\, nodes & $\gamma-1$ & $(t-t_{0})^{2/{3{\gamma}}}$\\
\hline

(2a),(2b) & $\pm(1+2 x_4^{\,\,2})^{1/2}$ &  $0$ &  $x_{4}$ &
{stable}\, attractor & $-1$ & $e^{\Lambda(t-
t_{0})}\,\,(\Lambda=constant)$\\ \hline\mr\ns
\end{tabular}
\end{center}
\end{table*}

\begin{figure}[htp]
\includegraphics{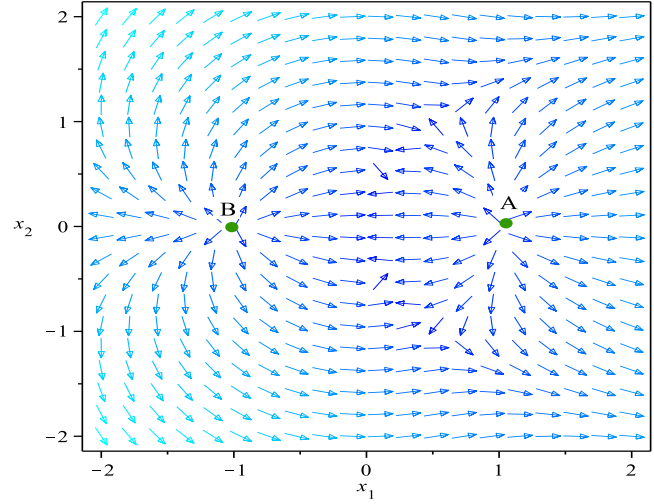} \vspace{8cm} \caption{\label{f1}\small { The phase
plane for $\lambda=+1$\,. Points $A$ and $B$ are unstable nods.}}
\end{figure}

The ingredients of \tref{t1} can be explained with more geometrical
details through the phase space trajectories. \Fref{f1} shows the
two dimensional $(x_{1}-x_{2})$ phase plane for $\lambda=+1$\,. In
this figure, points $A$ and $B$ both are unstable nodes.

\begin{figure}[htp]
\begin{center}\includegraphics{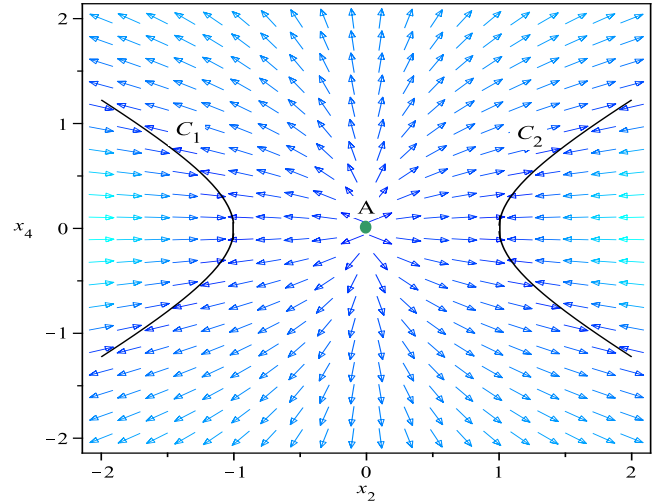} \vspace{7cm}
\end{center}
\caption{\label{f2}\small { The phase plane for  $w_{m}=0$. Point
$A$ is an unstable node(which reflects the first line of \tref{t2}).
There are two curves $(C_{1}$ and $C_{2})$, which are corresponding
to critical lines $((2a),(2b))$ of \tref{t2}.}}
\end{figure}

We note that the above arguments were based on the assumption that
$x_{4}=0$ for all critical points. Since $x_{4}$ is related directly
to the braneworld nature of the solutions, our analysis up to this
point was effectively 4-dimensional. Now we consider the case that
$x_{4}\neq 0$. \Tref{t2} gives the results of the corresponding
phase space analysis. In this case, there is a cosmological constant
dominated accelerating phase which is an stable attractor
corresponding to critical lines $(2a),(2b)$ of \tref{t2} (curves
$C_{1}$ and $C_{2}$ of \fref{f2}).

As we have stated previously, the pure normal DGP solution has not
the potential to explain the late-time acceleration. In our case
with a quintessence field on the brane, as we have shown, it is
possible to realize the late-time acceleration in this setup. We
note also that as \tref{t1} shows, for fixed points \{(1a),(1b)\},
\{(3a),(3b)\} and also \{(4a),(4b)\} it is possible in principle to
realize a late-time acceleration which depends on parameters
$\gamma$ and $\lambda$ and the stability of which needs to be
investigated in each case. For fixed points \{(1a),(1b)\} (first row
of \tref{t1}), the accelerating phase (with $q<0$ where $q\equiv
-\frac{a\ddot{a}}{{\dot{a}}^{2}}$) is possible if $\gamma < 2/3$.
But this accelerating phase is a saddle point and obviously is not
the late-time cosmic accelerating phase. For fixed point
\{(3a),(3b)\}, the accelerating phase is possible if
$\lambda^{2}<2$. This gives also a saddle point which is not
corresponding to the late-time stable, accelerating phase of cosmic
expansion. For fixed point \{(4a), (4b), (4c), (4d)\}, there is an
accelerating, stable phase if $\gamma < 2/3$. However, these stable
points are either a node or a spiral. Albeit these are not
corresponding to a late-time accelerating, stable, de Sitter
attractor. So, none of the accelerating phases corresponding to
\tref{t1} are the de Sitter phase. Nevertheless, as we have stated
previously, point (2) (actually a critical line) of \tref{t2} gives
a de Sitter attractor. Therefore, with a quintessence field in the
normal DGP setup, it is possible to realize the late-time
acceleration in contrast to pure DGP case.

\begin{figure}[htp]
\begin{center}\includegraphics{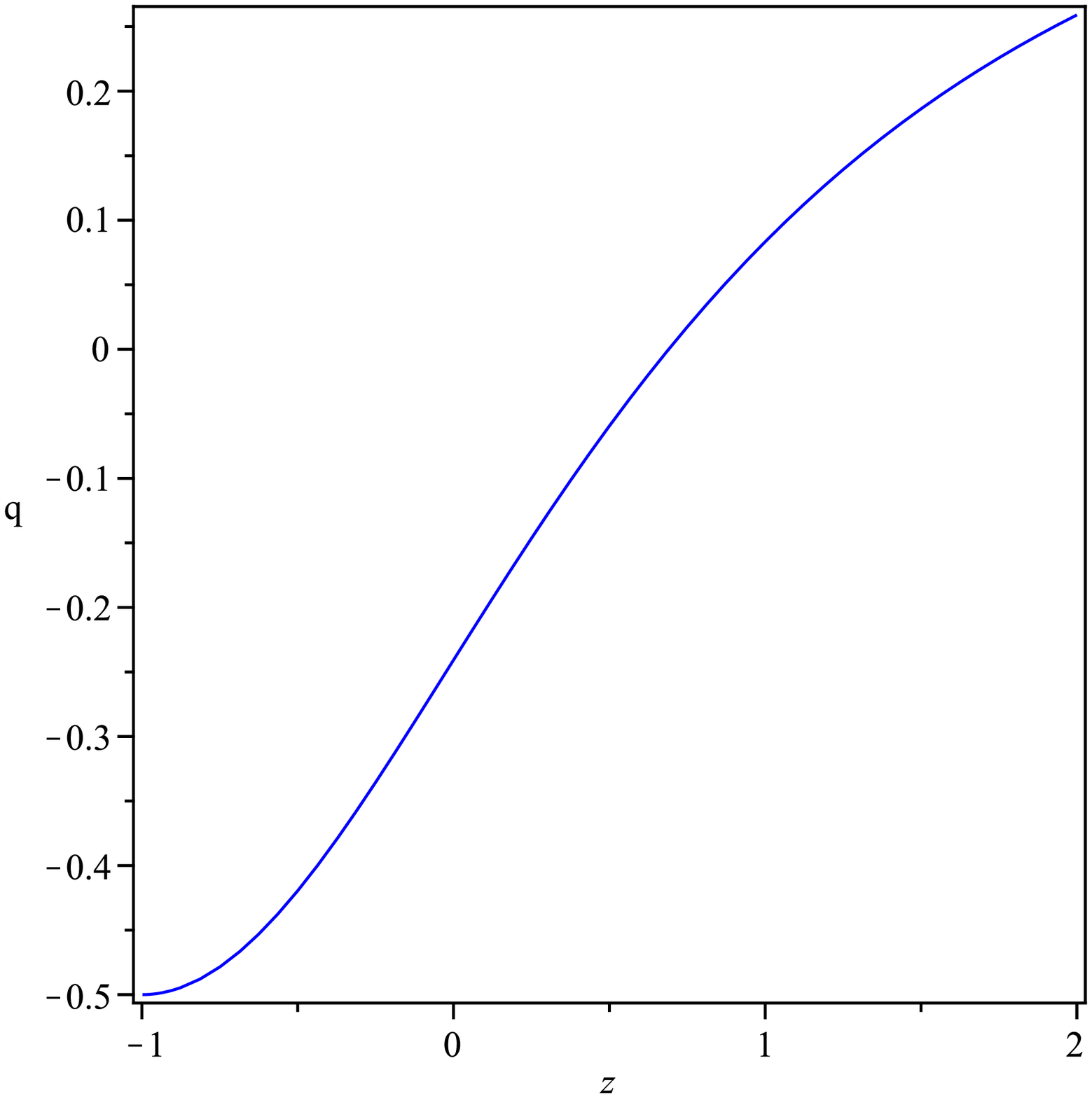} \vspace{7.2cm}
\end{center}
\caption{\label{f3}\small { The deceleration parameter versus the
redshift for $\Omega_{m}=0.28$, $\Omega_{\varphi}=0.8$, $w_{m}=0$
and $w_{\varphi}=-2/3$. Transition to the accelerating phase
occurs at $z=0.68$.}}
\end{figure}

Now we analyze the accelerating phase of the model through the
evolution of the deceleration parameter. The equation for the
deceleration parameter is $q=-\frac{\ddot {a} a}{{\dot
{a}}^{2}}=\frac{1}{2} (1+3 w_{eff})$. In our case, it can be written
by using the phenomenological parameters $\Omega_{m}$ and
$\Omega_{\varphi}$ as
\begin{equation}
q=\frac{1}{2}\,{\frac { \left( 1+3\,w_{{m}} \right) \Omega_{{m}}
\left( 1+z
 \right) ^{3(1+\,w_{{m}})}+ \left( 1+3\,w_{{\varphi }} \right) \Omega_{{
\varphi }} \left( 1+z \right) ^{3(1+\,w_{{\varphi
})}}}{\Omega_{{m}}
 \left( 1+z \right) ^{3(1+\,w_{{m}})}+\Omega_{{\varphi }} \left( 1+z
 \right) ^{3(1+\,w_{{\varphi }})}}}\,.\label{eq21}
\end{equation}
\Fref{f3} shows the behavior of $q$ versus the redshift. There is a
transition to the accelerating phase at $z=0.68$.

\begin{figure}[htp]
\begin{center}\includegraphics{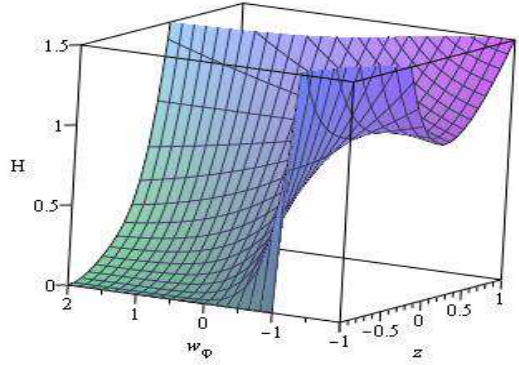} \vspace{6.2cm}
\end{center}
\caption{\label{f014}\small { The $3$-dimensional plot of the
Hubble parameter versus the redshift and equation of state
parameter of the scalar field. }}
\end{figure}

\begin{figure}[htp]
\begin{center}\includegraphics{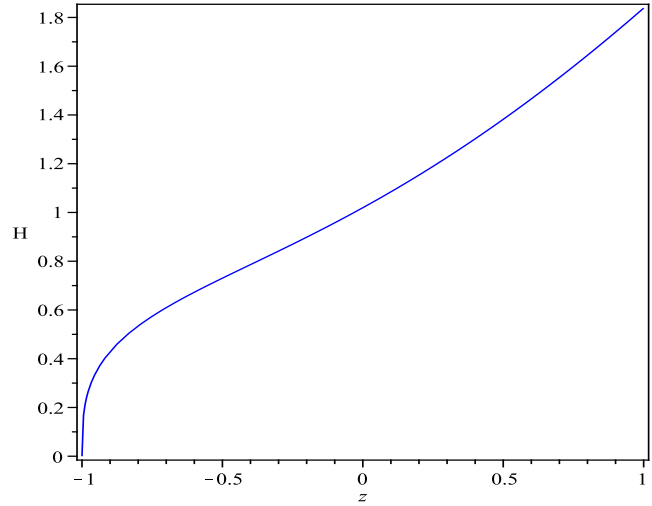} \vspace{7cm}
\end{center}
\caption{\label{f015}\small { This figure shows the plot of the
Hubble parameter versus the redshift for a quintessence field. It is
plotted for $\Omega_{m}=0.28$, $\Omega_{\varphi}=0.8$, $w_{m}=0$ and
$w_{\varphi}=-0.8$.}}
\end{figure}

We define the dimensionless density parameters as follows
$$\Omega_{m}=\frac{\rho_{0}}{3 H_{0}^{\,\,2}}, \qquad \Omega_{\varphi}=\frac{\rho_{\varphi 0}}{3 H_{0}^{\,\,2}},
 \qquad \Omega_{r_{c}}=\frac{1}{4 r_{c}^2 H_{0}^{\,\,2}}\,. $$
Then the Friedmann equation can be rewritten as
\begin{equation}
H(z)= H_{0}\, \left( \sqrt {\Omega_{{r_{{c}}}}+\Omega_{{m}} \left(
1+z
 \right) ^{3(1+\,w_{{m}})}+\Omega_{{\varphi }} \left( 1+z \right) ^{3(1+
\,w_{{\varphi }})}}-\sqrt {\Omega_{{r_{{c}}}}} \right)
\end{equation}
\Fref{f014} shows the evolution of the Hubble parameter versus the
redshift and equation of state parameter of a minimally coupled
quintessence scalar field with $\Omega_{m}=0.28$,
$\Omega_{\varphi}=0.8$. We can understand from this figure that the
Hubble parameter of the model decreases as the redshift decreases.
This is a trace of essentially possible realization of an
\emph{effective} phantom-like behavior on the brane (see \cite{11}
for more details). \Fref{f015} is a $2$-dimensional plot of $H$
versus the redshift for a quintessence field with $w_{\varphi}=-
0.8$.

\begin{figure}[htp]
\begin{center}\includegraphics{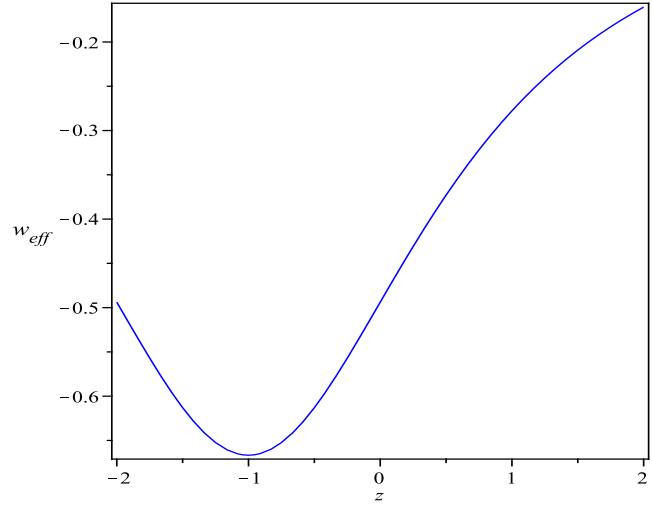} \vspace{7.2cm}
\end{center}
\caption{\label{f4}\small { The effective equation of state
parameter versus the redshift for $\Omega_{m}=0.28$,
$\Omega_{\varphi}=0.8$, $w_{m}=0$ and $w_{\varphi}=-2/3$. There is
no crossing of the cosmological constant line. }}
\end{figure}

Another important issue is the possibility of crossing of the
cosmological constant line (the so-called phantom-divide line) by
the equation of state parameter. The effective equation of state
parameter can be written as follows
\begin{equation}
w_{{{\it eff}}}={\frac {w_{{m}}\Omega_{{m}} \left( 1+z \right)
^{3(1+\, w_{{m}})}+w_{{\varphi }}\Omega_{{\varphi }} \left( 1+z
\right) ^{3(1+\,w _{{\varphi }})}}{\Omega_{{m}} \left( 1+z \right)
^{3(1+\,w_{{m}})}+\Omega _{{\varphi }} \left( 1+z \right)
^{3(1+\,w_{{\varphi }})}}}\,.\label{eq22}
\end{equation}
As \fref{f4} shows, it is impossible to cross the cosmological
constant equation of state parameter $w_{\varphi}=-1$ by a
minimally coupled quintessence field in the normal DGP setup.

\begin{figure}[htp]
\begin{center}\includegraphics{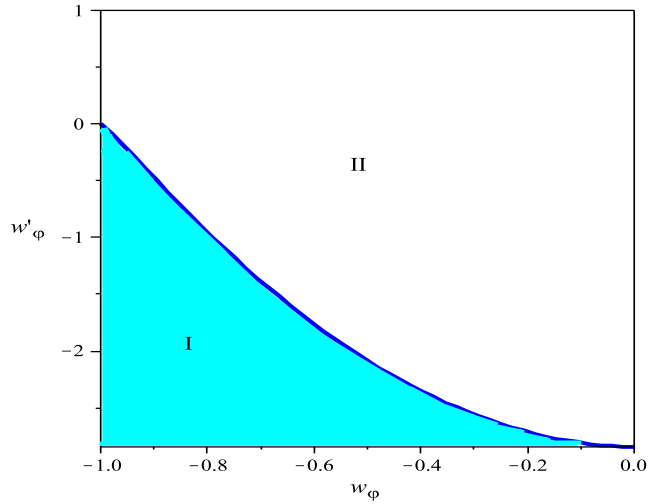} \vspace{7.2cm}
\end{center}
\caption{\label{f5}\small {Bounds on $w_{\varphi}^{\prime}$ as a
function of $w_{\varphi}$ in $w_{\varphi}-w'_{\varphi}$
phase-plane for $\Omega_{\varphi}=0.8$} and $\lambda=0.1$}
\end{figure}

Now we focus on the classical stability of the solutions in\,
$w_{\varphi}-w_{\varphi}'$\,  phase-plane of the present model (see
[15] for a similar analysis for other interesting cases). Here a
prime marks the derivative of $w_{\varphi}$ with respect to the
logarithm of the scale factor, $N=\ln a(t)$, so that
\begin{equation}
w'_{\varphi}\equiv\frac{dw_{\varphi}}{dN}={\frac{dw_{\varphi}}
{d\rho_{\varphi}}}{\frac{d\rho_{\varphi}}{dN}}\,.\label{eq23}
\end{equation}
We define the function $c_{a}$ so that
$c_{a}^{2}\equiv\frac{\dot{p}_{\varphi}}{\dot{\rho}_{\varphi}}$\,
or equivalently\,  $ c_{a}^{2}\equiv
\frac{dp_{\varphi}}{d\rho_{\varphi}}$.\, Generally the sound speed
expresses the phase velocity of the inhomogeneous perturbations of
the scalar field. If we suppose the scalar field's energy-momentum
to have a perfect fluid form, this function would be the adiabatic
sound speed of this fluid. To avoid the future big rip
singularity, we set $c^{2}_{a}>0$. Since
\begin{equation}
\eqalign{{\frac{dw_{\varphi}}{d\rho_{\varphi}}}&={\frac{1}{\rho_{\varphi}}}
{\frac{dp_{\varphi}}{d\rho_{\varphi}}}
-{\frac{p_{\varphi}}{\rho_{\varphi}^{\,\,2}}} \nonumber\\
 &={\frac{1}{\rho_{\varphi}}}\Big({\frac{dp_
{\varphi}}{d\rho_{\varphi}}}-w_{\varphi}\Big)}\,,\label{eq24}
\end{equation}
and
\begin{equation}
{\frac{d\rho_{\varphi}}{dN}}={\frac{\dot{\rho_{\varphi}}}{H}}=-3(1+w_{\varphi})\rho_{\varphi}\,,\label{eq25}
\end{equation}
we obtain
\begin{equation}
w'_{\varphi}=-3(1+w_{\varphi})\Big(c_{a}^{2}-w_{\varphi}\Big)\,.\label{eq26}
\end{equation}
Therefore, we obtain
\begin{equation}
w_{\varphi}^\prime=-3(1-w_{\varphi}^{\,\,2})+{\lambda}
{\sqrt{3(1+w_{\varphi})\Omega_{\varphi}}}(1-w_{\varphi})\,.\label{eq27}
\end{equation}
Now the $w_{\varphi}-w'_{\varphi}$ phase plane is divided into the
following two regions
\begin{equation}
\left\{\begin{array}{ll}w_{\varphi}>-1 \,\quad, w'_{\varphi}<3w_{\varphi}(1+w_{\varphi})&\quad \quad c_{a}^{2}>0\,\quad(region\, I)\\ \\
w_{\varphi}>-1
\,\quad,w'_{\varphi}>3w_{\varphi}(1+w_{\varphi})&\quad\quad{\rm
c_{a}^{2}<0}\,\quad (region\, II)\end{array}\right.\label{eq28}
\end{equation}
that are shown in \fref{f5}. The region I is the classical stability
region of the theory. We note that there is no phantom phase in this
case and therefore we have not encounter with four distinct regions
of $w_{\varphi}-w'_{\varphi}$ plane as usually are discussed in
literature.

\subsection{Minimally coupled phantom field on the normal DGP setup}

Astrophysical data indicate that $w$, the equation of state
parameter of the cosmic fluid, lies in a very narrow strip close to
$-1$. The case $w=-1$ corresponds to the cosmological constant. For
$w$ less than $-1$, the \emph{phantom} dark energy is observed, and
for $w$ more than $-1$ (but less than$-\frac{1}{3}$) the dark energy
is described by a quintessence field which has been studied in the
previous subsection. Moreover, the analysis of the properties of
dark energy from recent observational data mildly favor models of
dark energy with $w$ crossing the $-1$ line in the near past. So,
the phantom phase equation of state with $w < -1$ is mildly allowed
by observations \cite{16,17,18}. In this case, the universe
currently lives in its phantom phase which ends eventually at a
future singularity (the Big Rip singularity \cite{19}). There are
also a lot of evidence all around of a dynamical equation of state,
which has crossed the so called the phantom divide line $w=-1$
recently, at the value of redshift parameter $z\approx 0.25$
\cite{16,17,18}. For these reasons, now we pay our attention to a
phantom field localized on the DGP brane and we study cosmological
dynamics of the normal DGP branch in this case within a phase space
analysis.\\

The energy density and pressure of a phantom field are defined as
\begin{equation}
\rho_{\varphi}=-\frac{1}{2} {\dot{\varphi}}^2
+V(\varphi)\,,\label{eq29}
\end{equation}
\begin{equation}
p_{\varphi}=-\frac{1}{2} {\dot{\varphi}}^2
-V(\varphi)\,,\label{eq30}
\end{equation}
respectively. The Klein-Gordon equation governing on the dynamics
of the phantom field is given by
\begin{equation}
\ddot{\varphi}+3H{\dot{\varphi}}=\frac{dV}{d\varphi}\,.\label{eq31}
\end{equation}

With phase space coordinates as defined in \eref{eq8}, The Friedmann
equation of the model in phase space now is written as

\begin{equation}
-x_1^2+x_2^2+x_3^2-2x_4^2=1\,,\label{eq32}
\end{equation}
and the autonomous dynamical equations are

\begin{equation}
\frac{dx_1}{dN}=-3x_1-{\frac{\sqrt{6}}{2}}\lambda x_2^2+3x_1
\Bigg({\frac{-2x_{1}^{\,\,2}+(1+w)x_{3}^{\,\,2}}{1-x_{1}^{\,\,2}
+x_{2}^{\,\,2}+x_{3}^{\,\,2}}}\Bigg)\,,\label{eq33}
\end{equation}

\begin{equation}
\frac{dx_2}{dN}=-\frac{\sqrt{6}}{2}\lambda x_1
x_2+3x_2\Bigg({\frac{-2x_{1}^{\,\,2}+(1+w)x_{3}^{\,\,2}}
{1-x_{1}^{\,\,2}+x_{2}^{\,\,2}+x_{3}^{\,\,2}}}\Bigg)\,,\label{eq34}
\end{equation}

\begin{equation}
\frac{dx_{3}}{dN}=-\frac{3}{2}x_3\Bigg({\frac{-2x_{1}^{\,\,2}
+(1+w)x_{3}^{\,\,2}}{1-x_{1}^{\,\,2}+x_{2}^{\,\,2}+x_{3}^{\,\,2}}}\Bigg)\,.\label{eq35}
\end{equation}

In this case there are eight critical points that are shown in
\tref{t3}. The eigenvalues of these points are

$\bullet$ points (1a) , (1b):
$$\alpha_{1}=\frac{3}{2}(\gamma-2),\quad
\alpha_{2}=\frac{3}{2}\gamma,\quad \alpha_{3}=\frac{3}{2}\gamma$$

$\bullet$ points (2a) , (2b):
$$\alpha_{1}=-\frac{1}{2}(\lambda^2+6),\quad
\alpha_{2}=-\frac{1}{2}({\lambda^2+3\gamma}),\quad
\alpha_{3}=-\frac{\lambda^2}{2}$$

$\bullet$ points (3a) , (3b) , (3c) , (3d):
$$\alpha_{1}=\frac{3}{2}\gamma,\quad \alpha_{2,3}=-\frac{3}{4}(2-\gamma)
\Bigg(1\pm{\sqrt{1-\frac{8\gamma(\lambda^2+3\gamma)}{\lambda^2(2-\gamma)}}}\Bigg)\,.$$

\begin{table*}
\begin{tiny}
\begin{center}
\caption{\label{t3}\small{Location and dynamical character of the
fixed points.}}
\begin{tabular}{|c|c|c|c|c|c|c|c|c|c|}\hline\mr\ns

 name & $x_{1c}$ & $x_{2c}$ & $x_{3c}$ & Existence  &
stability &
$\Omega_{\varphi}$ &$ \gamma_{\varphi}$&$w_{eff}$&$a(t)$\\
\hline\mr\ns

(1a),(1b)& 0 & 0 & $\pm{1}$ &$\forall \, \lambda,\, \gamma$ &
saddle point
& 0 & undefined & $\gamma-1$ & $a_{0}(t-t_{0})^{2/{3\gamma}}$\\
\hline

(2a),(2b)&
$-{\frac{\lambda}{\sqrt{6}}}$&$\pm(1+{\frac{\lambda^2}{6}})^{1/2}$
& 0 & $\forall \, \lambda,\, \gamma$ & stable node  & $1$
&$\frac{-\lambda^2}{3}$
& $\frac{-\lambda^2}{3}-1$ &$a_{0}(t-t_{0})^{-2/\lambda^{2}}$ \\
\hline

(3a),(3b)&&&&&&&&&\\
(3c),(3d)& ${\sqrt{\frac{3}{2}}}{\frac{\gamma}{\lambda}}$&
$\pm({\frac{3\gamma(\gamma-2)}{2\lambda^2}})^{1/2}$ &
$\pm{(1+\frac{3\gamma}{\lambda^2})^{1/2}}$ &
$\gamma<0\,\,,\,\,\lambda^2>-3\gamma$& saddle point &$
{-\frac{3\gamma}{\lambda^2}}$&$\gamma $ & $\gamma-1$
&$a_{0}(t-t_{0})^{\frac{2}{3\gamma}}$\\ \hline\mr\ns
\end{tabular}
\end{center}
\end{tiny}
\end{table*}

In \tref{t3} we summarized also the results of the phase space
analysis of the model in addition to cosmological characters of each
critical point. For critical points $\{(1a),(1b)\}$, there is no
contribution of the phantom scalar field and the universe is
dominated by matter fields other than the phantom scalar field.
These points behave like saddle points in the phase space and for
$\gamma<2/3$, these points give an accelerating phase. It is
possible to have scaling solutions in this case too. For points
$\{(2a),(2b)\}$, there is no contribution of ordinary matter fields
and the energy density of the universe is dominated by the phantom
scalar field's kinetic and potential energies. In these cases, there
is no possibility of accelerated expansion on the brane. The
corresponding points in phase plane are stable nodes. The last line
of \tref{t3} consists of four critical points
$\{(3a),(3b),(3c),(3d)\}$. These are just saddle points in phase
plane. There are scaling solutions in these critical points. For
these fixed points accelerated expansion is possible if
$\gamma<2/3$. However, this accelerated expansion phase is not a de
Sitter stable phase.\\

\begin{figure}[htp]
\begin{center}\includegraphics{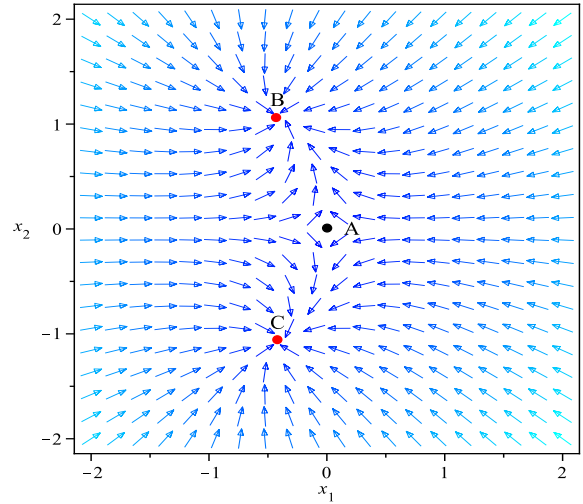} \vspace{7cm}
\end{center}
\caption{\label{f6}\small { The phase plane for $\lambda=+1$\, and
\,$w_{m}=0 $. Point A is a saddle point, whereas points B and C are
stable nodes .}}
\end{figure}

\begin{figure}[htp]
\begin{center}\includegraphics{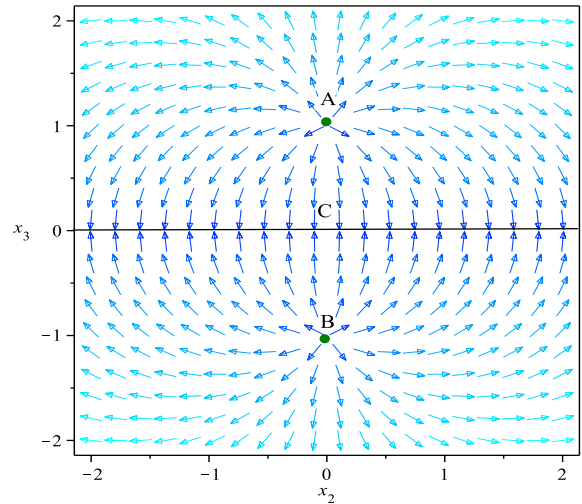} \vspace{7cm}
\end{center}
\caption{\label{f7}\small { The $x_{3}-x_{2}$ phase plane for
$w_{m}=0 $. Points $A$  and $B$ are unstable nodes (which reflects
the first row of table 2). There exists a critical line in this
case.}}
\end{figure}

We note that the analysis presented in the previous paragraph was
based on the condition $x_{4}=0$ which gives essentially an
effective 4-dimensional picture of the model. Now we consider the
case that $x_{4}\neq 0$. The phase space analysis of the model with
minimally coupled phantom field gives \emph{the same} results as are
presented in \tref{t2} for a minimally coupled quintessence field.
Similar to minimal quintessence scalar field case, in this case
there is a cosmological constant dominated accelerating phase which
is an stable attractor corresponding to critical lines $(2a),(2b)$
of \tref{t2} (curves $C_{1}$ and $C_{2}$ of \fref{f2}). So, with a
phantom field minimally coupled to induced gravity in the normal DGP
setup it is possible to have an attractor, de Sitter solution
realizing the late-time accelerated expansion (see \tref{t2}).
\Fref{f6} shows a plot of the $x_{1}-x_{2}$ phase plane of the
model. Point $A$ is corresponding to points (1a) and (1b) of
\tref{t3}. Points $B$ and $C$ are corresponding to points (2a) and
(2b) of \tref{t3} which are stable nodes. The $x_{3}-x_{2}$ (with
$x_{4}\neq 0$) phase plane of the model is shown in \fref{f7}. We
note that this figure is actually the same as \fref{f2} but now
plotted in $x_{3}-x_{2}$ plane rather than $x_{2}-x_{4}$ plane. Here
we encounter a line of stability points as is shown in \fref{f7}.
Points $A$ and $B$ are corresponding to critical points $(1a)$ and
$(1b)$ of \tref{t2}.

\begin{figure}[htp]
\begin{center}\includegraphics{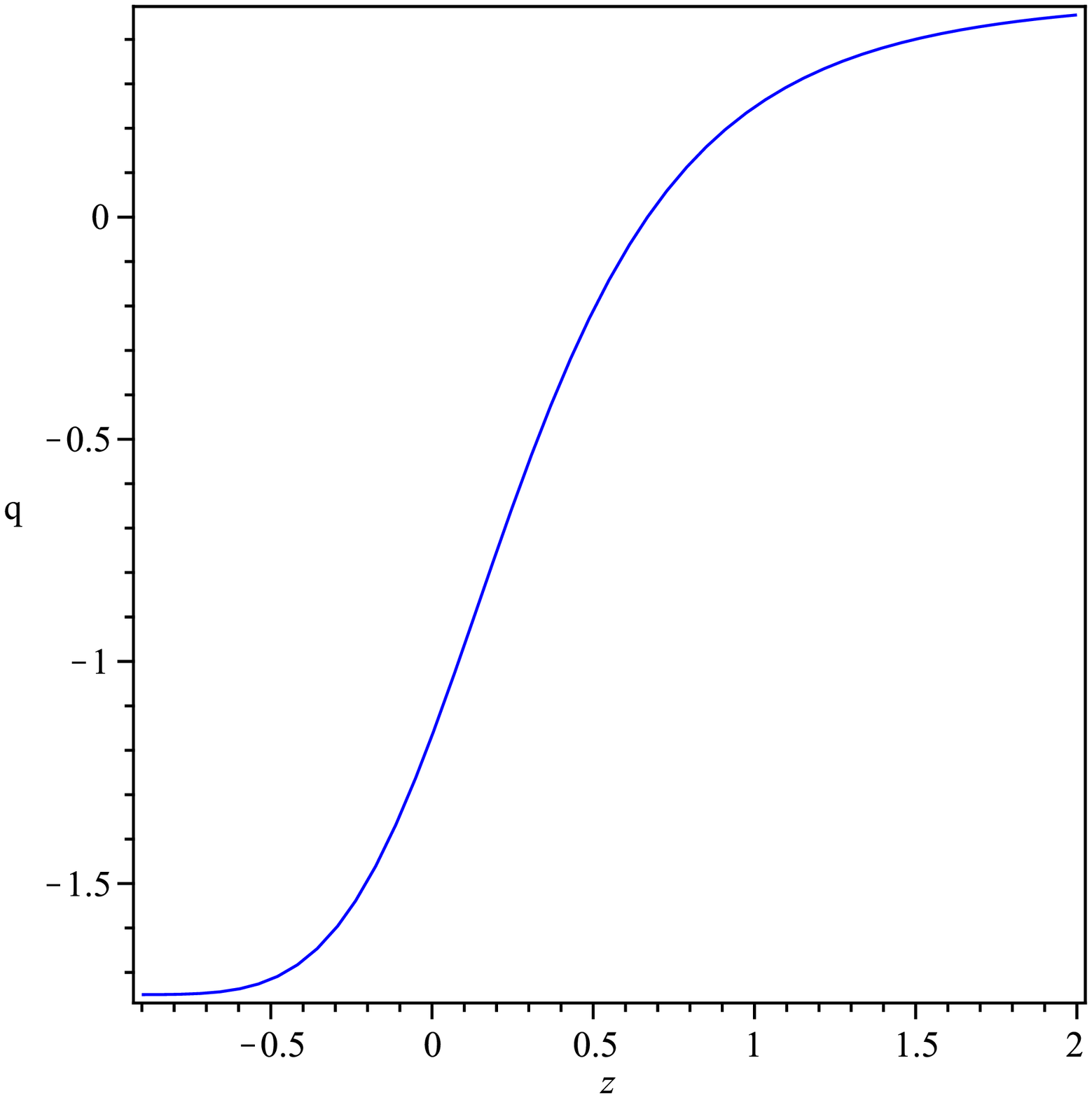} \vspace{7.5cm}
\end{center}
\caption{\label{f8}\small {The deceleration parameter versus the
redshift for $\Omega_{m}=0.28$, $\Omega_{\varphi}=0.8$, $w_{m}=0$
and $w_{\varphi}=-1.5$. Transition to the accelerating phase
occurs at $z=0.67$.}}
\end{figure}

\begin{figure}[htp]
\begin{center}\includegraphics{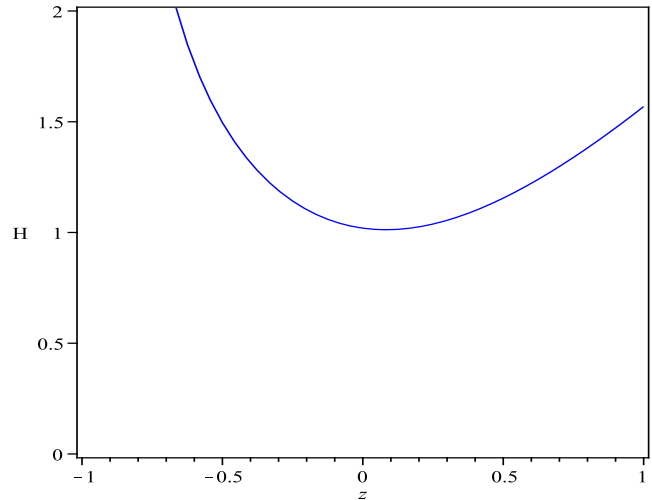} \vspace{7cm}
\end{center}
\caption{\label{f016}\small { Evolution of the Hubble parameter with
redshift on the brane with a phantom field. Here we considered
$w_{\varphi}=-1.5$.}}
\end{figure}

\begin{figure}[htp]
\begin{center}\includegraphics{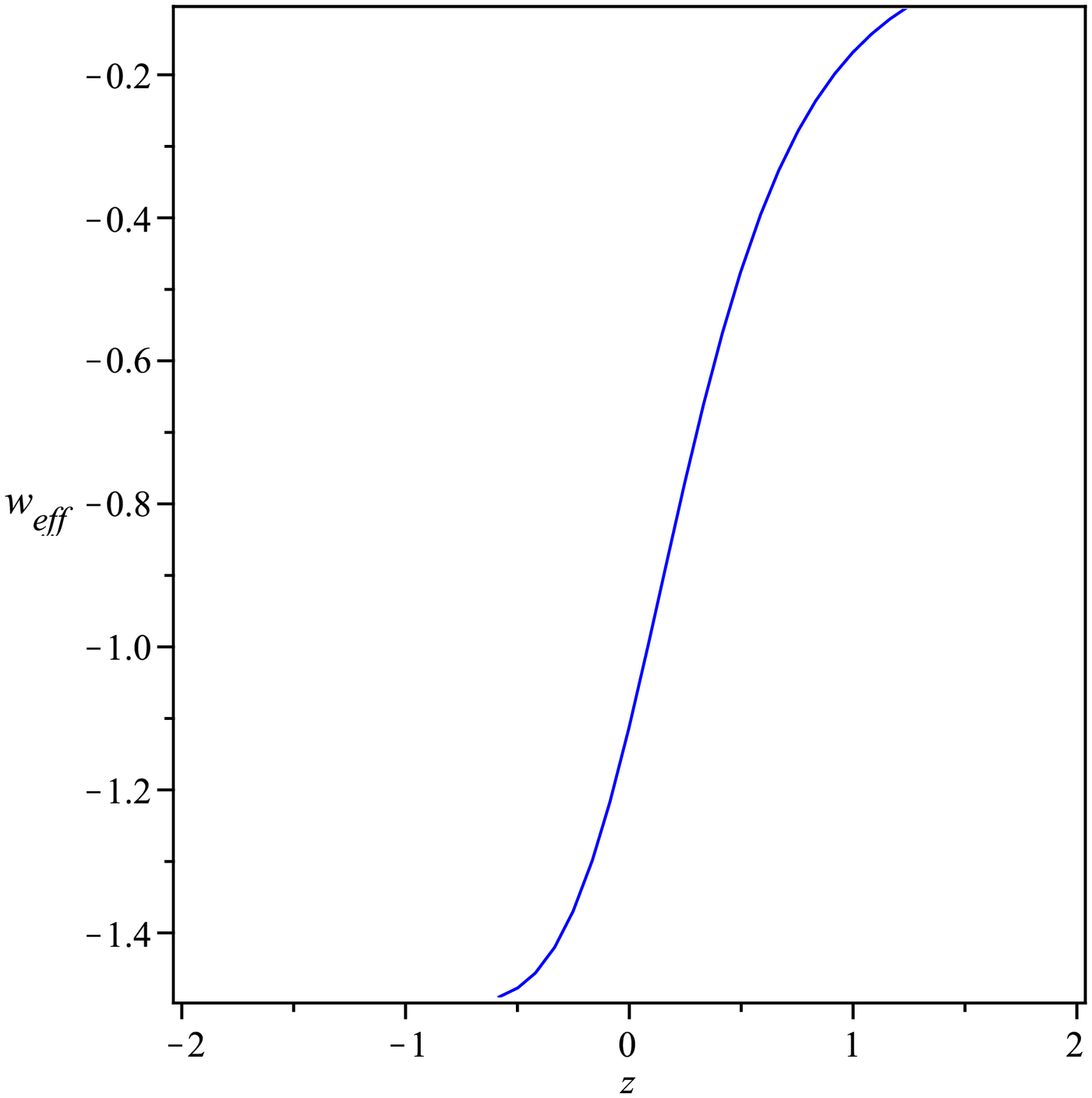} \vspace{7cm}
\end{center}
\caption{\label{f9}\small {The effective equation of state parameter
versus the redshift for $\Omega_{m}=0.28$, $\Omega_{\varphi}=0.8$,
$w_{m}=0$ and $w_{\varphi}=-1.5$. Transition to the phantom phase
occurs at $z\sim0.25$.}}
\end{figure}

The form of the deceleration parameter and effective equation of
state parameter for phantom field are the same as quintessence field
presented as equations (21) and (23). \Fref{f8} shows the behavior
of the deceleration parameter $q(z)$. The universe enters the
accelerated phase at $z\simeq 0.68$. Also \Fref{f016} shows how the
normal branch Hubble parameter evolves on the brane with a phantom
field. Also \Fref{f9} shows the behavior of $w_{eff}(z)$. There is a
crossing of the phantom divide line in this setup. Therefore, with a
minimally coupled phantom field on the normal DGP setup, it is
possible to cross the phantom divide line by the effective equation
of state parameter of the dark energy. Note that as we have shown
previously, this crossing was impossible with a quintessence field
on the brane.

To investigate the classical stability of the solutions with phantom
field in the normal DGP setup and within the
$w_{\varphi}-w'_{\varphi}$ phase-plane approach, we adopt the same
strategy as has been done for quintessence field. For this minimally
coupled phantom field, we have

\begin{equation}
w_{\varphi}^\prime=-3(1-w_{\varphi}^{\,\,2})+{\lambda}
{\sqrt{-3(1+w_{\varphi})\Omega_{\varphi}}}(1-w_{\varphi})\,.\label{eq36}
\end{equation}
Now the $w_{\varphi}-w'_{\varphi}$ phase-plane is divided into the
following two regions
\begin{equation}
\left\{\begin{array}{ll}w_{\varphi}<-1 \,,\quad w'_{\varphi}<3w_{\varphi}(1+w_{\varphi})&\quad \quad c_{a}^{2}>0\,\quad(region\, I)\\ \\
w_{\varphi}<-1 \,, \quad
w'_{\varphi}>3w_{\varphi}(1+w_{\varphi})&\quad\quad{\rm
c_{a}^{2}<0}\,\quad (region\, II)\end{array}\right.\label{eq37}
\end{equation}
The \emph{Region I} is the subspace of the classical stability of
the solutions with a minimally coupled phantom field on the brane.

\begin{figure}[htp]
\begin{center}\includegraphics{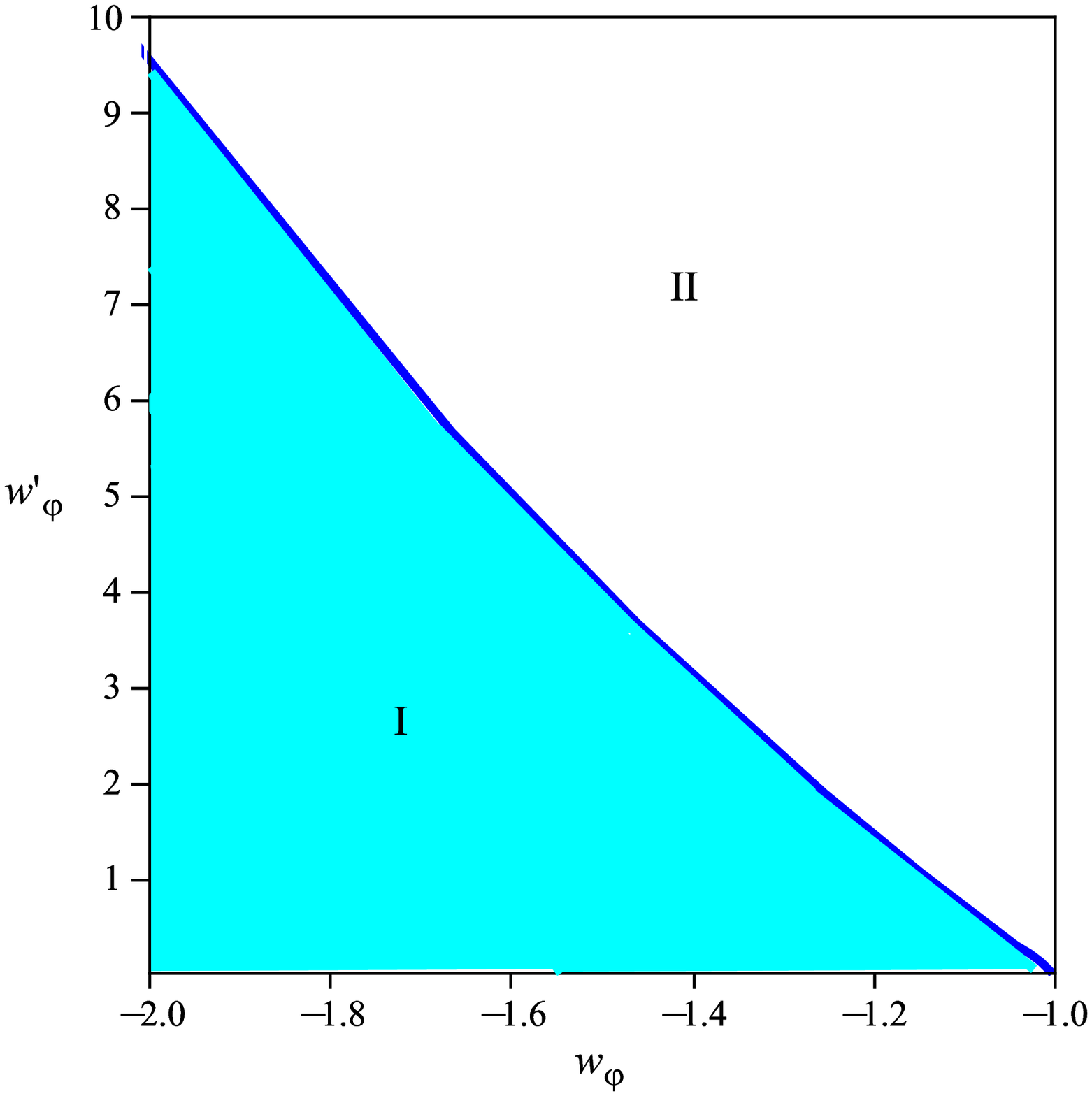} \vspace{7cm}
\end{center}
\caption{\label{f10}\small {Bounds on $w_{\varphi}^{\prime}$ as a
function of $w_{\varphi}$ in $w_{\varphi}-w'_{\varphi}$
phase-plane for $\Omega_{\varphi}=0.8$} and $\lambda=0.1$. The
subspace shown by \emph{Region I} of the model parameter space is
the classical stability subspace of the solutions. }
\end{figure}

\section{Cosmological dynamics of a non-minimally coupled scalar field on the normal DGP setup}

Now we consider a non-minimally coupled scalar field on the DGP
brane. As we have stated previously, in a realistic gravitational or
cosmological scenario with scalar fields, incorporation of the
non-minimal coupling is inevitable. In fact, incorporation of a
non-minimal coupling (NMC) between matter field and gravity is
necessary from several compelling reasons. There are many
theoretical evidences that suggest incorporation of an explicit
non-minimal coupling of the scalar field and gravity in the action
\cite{20,21}. A nonzero non-minimal coupling arises from quantum
corrections and it is required also for renormalizability of the
corresponding field theory. Amazingly, it has been proven that the
phantom divide line crossing of the dark energy described by a
single, minimally coupled scalar field with a general Lagrangian is
even unstable with respect to the cosmological perturbations
realized on the trajectories of the zero measure \cite{22}. This
fact has motivated a lot of attempts to realize crossing of the
phantom divide line by equation of state parameter of a scalar field
non-minimally coupled to gravity as dark energy candidate in more
complicated frameworks \cite{23}. In which follows, we study
cosmological dynamics with a non-minimally coupled
quintessence/phantom field on the normal DGP setup within a phase
space approach analysis. We study possible realization of the
late-time acceleration and crossing of the phantom divide line in
this setup. We also investigate the classical stability of the
solutions in separate regions of the $w-w'$ phase-plane.

\subsection{Non-minimally coupled quintessence field on the normal DGP branch}

The equations governing on the cosmological dynamics on the normal
DGP branch with a non-minimally coupled quintessence field are as
follows

\begin{equation}
\dot{H}=-{\frac{\rho+p}{2\mu^{2}}}\Big(1+{\frac{1}{2Hr_{c}}}\Big)^{-1}\,,\label{38}
\end{equation}
where $\rho=\rho_{m}+\rho_{\varphi}$ and $p=p_{m}+p_{\varphi}$\,,
and the energy density and pressure of the scalar field are defined
as \cite{7}
\begin{equation}
\rho_{\varphi}=\frac{1}{2}{\dot{\varphi}^{2}}+V(\varphi)+6\xi
H\varphi{\dot{\varphi}}+3\xi H^{2}{\varphi}^{2}\,,\label{39}
\end{equation}
and
\begin{equation}
p_{\varphi}=\frac{1}{2}{\dot{\varphi}^{2}}-V(\varphi)-2\xi
(\varphi{\ddot{\varphi}}+2\varphi
H{\dot{\varphi}}+{\dot{\varphi}^{2}})-\xi
{\varphi}^{2}(2\dot{H}+3H^{2})\,,\label{40}
\end{equation}
respectively. As usual, $\rho_{m}$ is the energy density of
ordinary matter fields other than the scalar field $\varphi$ on
the brane. The other dynamical equation is the following
Klein-Gordon equation

\begin{equation}
\ddot{\varphi}+3H{\dot{\varphi}}+\xi R
\varphi=-{\frac{dV}{d\varphi}}\,.\label{41}
\end{equation}

Now we define the dimensionless variables as

\begin{equation}
\eqalign{x_1={\frac{\dot{\varphi}}{\sqrt{6}\mu H}}\,, \qquad
x_2={\frac{\sqrt{V}}{\sqrt{3}\mu H}}\,, \qquad
x_3={\frac{\sqrt{\rho_m}}{\sqrt{3}\mu H}}\,,\\
x_4={\frac{1}{\sqrt{2r_c H}}}\,, \qquad
x_{5}=\frac{\sqrt{\xi}}{\mu} \varphi\,.} \label{42}
\end{equation}

By using these phase space variables, now the evolution equations
\eref{38} and \eref{41} can be rewritten as
\begin{equation}
\fl \dot{H}=-{\frac{3(1-2\xi)x_{1}^{\,\,2}+\sqrt{6\xi}x_{1}
x_{5}+{\frac{3}{2}}\gamma
x_{3}^{\,\,2}+3\sqrt{\xi}x_{5}(\sqrt{6}x_{1}+4\sqrt{\xi}x_{5}-\lambda
x_{2}^{\,\,2})}{1+x_{4}^{\,\,2}-(1-6\xi)x_{5}^{\,\,2}}}H^{2}\,,\label{43}
\end{equation}
and
\begin{equation}
\eqalign{\ddot{\varphi}=-3\mu H^{2} \Bigg [&
\frac{(-2\sqrt{\xi}x_{5})\Big(3(1-2\xi)x_{1}^{\,\,2}+\sqrt{6\xi}x_{1}x_{5}+{\frac{3}{2}}
\gamma x_{3}^{\,\,2}\Big)}{1+x_4^{\,\,2}-(1-6\xi)x_{5}^{\,\,2}} \nonumber\\
&+{\frac{(1+x_{4}^{\,\,2}-x_{5}^{\,\,2})(\sqrt{6}x_{1}+4\sqrt{\xi}x_{5}-\lambda
x_{2}^{\,\,2})}{1+x_4^{\,\,2}-(1-6\xi)x_{5}^{\,\,2}}}\Bigg]\,,}\label{44}
\end{equation}
respectively. Therefore, we obtain the following Friedmann
constraint equation in the phase space of the model
\begin{equation}
x_{1}^{\,\,2}+x_{2}^{\,\,2}+x_{3}^{\,\,2}+2\sqrt{6\xi}x_{1}x_{5}+x_{5}^{\,\,2}-2x_{4}^{\,\,2}=1\,.\label{45}
\end{equation}
To describe the dynamical system of the model, first we need to
obtain the autonomous, phase space equations. We differentiate the
phase space dimensionless variables with respect to $N=\ln{a}$ to
find
\begin{equation}
\frac{dx_1}{dN}=-3x_1+\frac{\sqrt{6}}{2}\lambda
x_2^{\,\,2}-2\sqrt{6\xi}x_5-\Big(\sqrt{6\xi}x_{5}+x_1\Big)\Psi\,,\label{46}
\end{equation}
\begin{equation}
\frac{dx_2}{dN}=-\frac{\sqrt{6}}{2}\lambda x_{1}
x_{2}-x_{2}\Psi\,,\label{47}
\end{equation}
\begin{equation}
\frac{dx_4}{dN}=-{\frac{1}{2}}x_{4}\Psi\,,\label{48}
\end{equation}
and
\begin{equation}
\frac{dx_5}{dN}=\sqrt{6\xi}x_{1}\,,\label{49}
\end{equation}
where by definition,
$$
\Psi\equiv\Bigg[-{\frac{3(1-2\xi)x_{1}^{\,\,2}+\sqrt{6\xi}x_{1}
x_{5}+{\frac{3}{2}}\gamma
x_{3}^{\,\,2}+3\sqrt{\xi}x_{5}(\sqrt{6}x_{1}+4\sqrt{\xi}x_{5}-\lambda
x_{2}^{\,\,2})}{1+x_{4}^{\,\,2}-(1-6\xi)x_{5}^{\,\,2}}}\Bigg]\,.$$

The stability around the fixed points is related to the form of the
eigenvalues in each critical point. The eigenvalues can be obtained
by using the above autonomous equations, the results of which are as
follows\\

$\bullet$ point (1a),(1b):$$\alpha_{1,2}=\frac{3}{2}\gamma,\quad
\alpha_{3,4}=-\frac{3}{2}+\frac{3}{4} \gamma\pm\frac{1}{4}
\sqrt{36-36 \gamma+9 \gamma^2-192 \xi+144 \xi \gamma}$$

$\bullet$ point (2a),(2b):$$\alpha_{1}=-\frac{3}{2}\gamma+2,\quad
\alpha_{2}=-1,\quad \alpha_{3,4}=2$$

$\bullet$ curve C:$$\alpha_{1}=-\frac{3}{2}\gamma,\quad
\alpha_{2}=0,$$

$$\alpha_{3,4}=\frac{1}{2}\frac{1}{(16\,\xi+16\,{x_{{2}}}^{2}\xi-{\lambda}^{2}{x_{{2}}}^{4}+12\,{\lambda}^{
2}{x_{{2}}}^{4}\xi)}\Bigg[-48\,\xi-48\,{x_{{2}}}^{2}\xi+3\,{\lambda}^{2}{x_{{2}}}^{4}-36\,{
\lambda}^{2}{x_{{2}}}^{4}\xi\,\,$$
$$\pm \Big(2304\,{\xi}^{2}+4608\,{x_{{2}}}^{2}{\xi}^{2}-288\,{\lambda}^{2}{x_{{2}
}}^{4}\xi+1920\,{\lambda}^{2}{x_{{2}}}^{4}{\xi}^{2}+2304\,{x_{{2}}}^{4
}{\xi}^{2}-288\,{x_{{2}}}^{6}\xi\,{\lambda}^{2}$$$$+4992\,{x_{{2}}}^{6}{
\xi}^{2}{\lambda}^{2}+9\,{\lambda}^{4}{x_{{2}}}^{8}-72\,{\lambda}^{4}{
x_{{2}}}^{8}\xi
+1872\,{\lambda}^{4}{x_{{2}}}^{8}{\xi}^{2}-24576\,{x_{{2}}}^{2}{\xi}^{3
}-12288\,{\xi}^{3}$$$$+384\,{\lambda}^{4}{x_{{2}}}^{6}\xi-3072\,{\lambda}^
{2}{x_{{2}}}^{2}{\xi}^{2}-12288\,{x_{{2}}}^{4}{\xi}^{3}-12\,{\lambda}^
{6}{x_{{2}}}^{10}-9216\,{x_{{2}}}^{6}{\xi}^{3}{\lambda}^{2}$$$$-9216\,{
\lambda}^{2}{x_{{2}}}^{4}{\xi}^{3}+144\,{\lambda}^{6}{x_{{2}}}^{10}\xi
-2304\,{\lambda}^{4}{x_{{2}}}^{6}{\xi}^{2}\Big)^\frac{1}{2}\Bigg]$$

We note that when there is a zero eigenvalue for a critical point,
it is necessary to use the \emph{center manifold theory} in order to
study the stability of that point in phase space of the model. In
our case there is a zero eigenvalue for a \emph{critical line}, and
therefore there is no need to do the center manifold analysis. In
other words, since we have a critical line here, the non-vanishing
eigenvalues are enough in this case to treat the stability of the
critical points (see for instance \cite{24}). \Tref{t4} summarizes
the results of the stability analysis in the phase space of this
model. Also this table contains types of possible cosmological
dynamics in this setup.

\begin{table*}
\begin{center}
\caption{\small{Location and dynamical character of the fixed
points.}} \label{t4}
\begin{tabular}{|c|c|c|c|c|c|c|c|c|c|} \hline\mr\ns

 name & $x_{1c}$ & $x_{2c}$&$x_{3c}$ &$x_{5c}$  & stability  &$ \gamma_{\varphi}$&$w_{eff}$&$a(t)$\\
\hline\mr\ns

(1a),(1b) & 0 & 0 & $\pm1$ & 0  & saddle point&undefined &
$\gamma-1$ & $a_{0}(t-t_{0})^{\frac{2}{3\gamma}}$\\
\hline

(2a),(2b) & 0 & 0 & 0 &$ \pm1$  & saddle point  & $4/3$ & $1/3$ &
$a_{0}(t-t_{0})^{\frac{1}{2}}$\\ \hline

C & 0 & $x_{2}$ & 0 & $\frac{\lambda x_{2}^{\,\,2}}{4 \sqrt{\xi}}$
& stable attractor& $0$ & $-1$ & $e^{\Lambda (t-t_{0})}$  \\
\hline\mr\ns
\end{tabular}
\end{center}
\end{table*}

\begin{figure}[htp]
\begin{center}\includegraphics{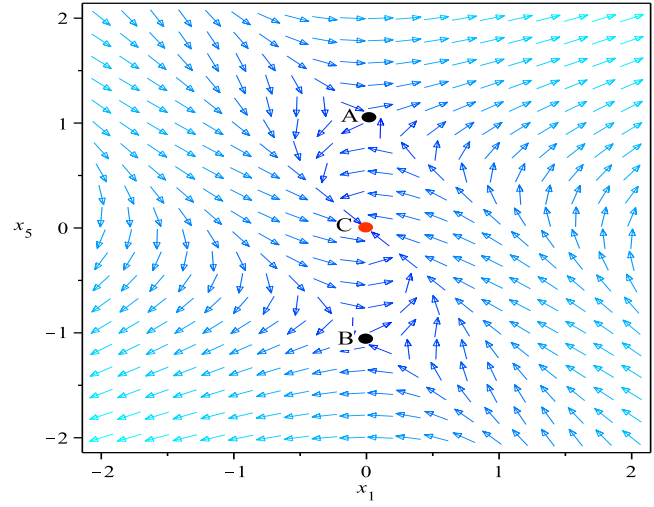} \vspace{7cm}
\end{center}
\caption{\small {\label{f11} The phase plane for $\xi=1/6$. Point
$C$ is a stable attractor, but points $A$ and $B$ are saddle
points.}}
\end{figure}

For critical points $\{(1a),(1b)\}$, there is no contribution of the
scalar field and the universe is dominated by matter fields other
than the quintessence scalar field. These two critical points behave
like saddle points in the phase space. For $\gamma <2/3$, one can
obtain an accelerating phase of expansion, but this phase is not
stable. It is possible to have scaling solutions for these cases. As
an interesting case, if $\gamma =2/3$, then we find $w_{eff}=-1/3$
which shows domination of the curvature energy. The critical points
$\{(2a),(2b)\}$ also behave like saddle points in the phase space
and in these cases we have no late-time acceleration. For these
points, the universe is radiation dominated. The last line of
\tref{t4}, stands for a critical line $(C)$ and in this case, there
is a cosmological constant dominated accelerating phase. In this
case, the potential energy of the brane scalar field plays the role
of a cosmological constant on the brane. So, with a non-minimally
coupled quintessence scalar field on the DGP brane, it is possible
to realize a stable, de Sitter late-time accelerating phase even in
the normal branch of the model. \Fref{f11} shows the phase plane of
the model with $\xi=1/6$ (the conformal coupling). Point $C$ is a
stable attractor, whereas points $A$ and $B$ are saddle points.\\

\begin{figure}[htp]
\begin{center}\includegraphics{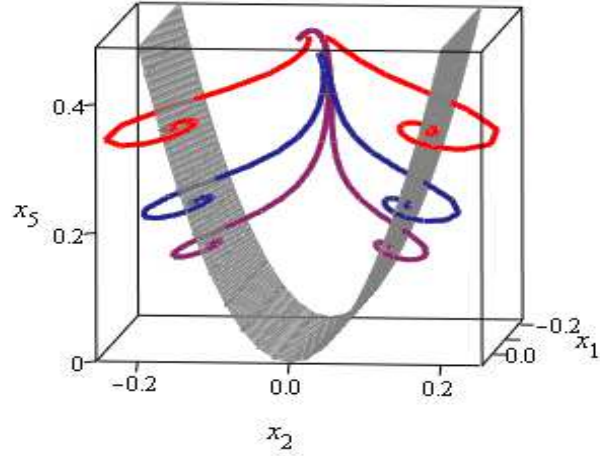} \vspace{7cm}
\end{center}
\caption{\small\label{f12} { The 3-dimensional phase plane for
$\xi=1/6$\,. The hyperbolic surface contains the attractor, stable
phases of the model.}}
\end{figure}

\Fref{f12} shows the three-dimensional, $x_{1}-x_{2}-x_{5}$ phase
space of the model for the last row of \tref{t4}. The stable,
attractor points are located in a hyperbolic curve (curve $C$ of
\tref{t4}) in $x_{2}-x_{5}$ perspective. In the three dimensional
$x_{1}-x_{2}-x_{5}$ phase space, this is a hyperbolic hypersurface
of stability points as shown in \fref{f12}. All points of this
hypersurface are stable attractors. We note that point $C$ of
\fref{f11} is corresponding to the mentioned hyperbolic on
$x_{1}-x_{5}$ plane.

\begin{figure}[htp]
\begin{center}\includegraphics{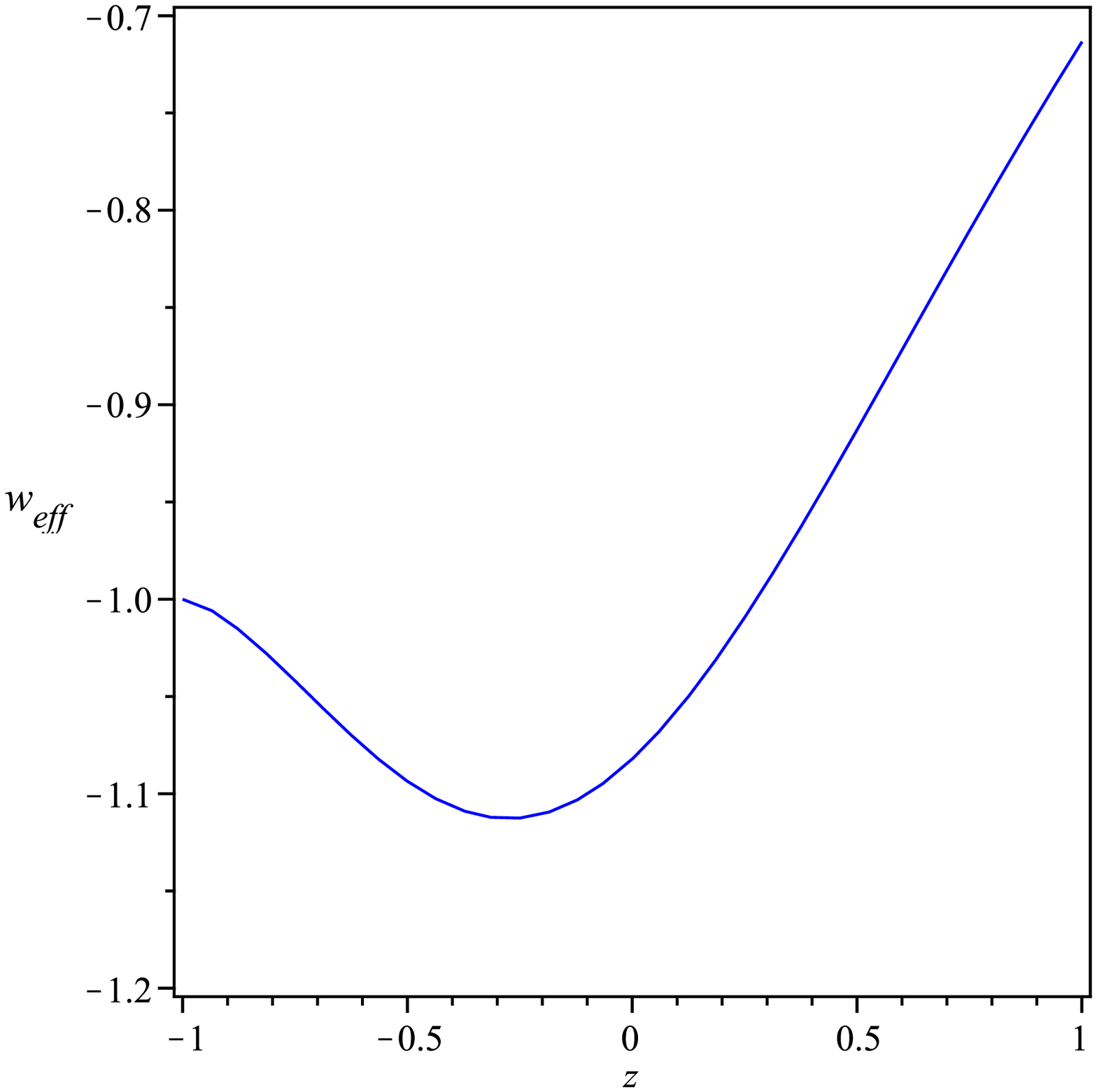} \vspace{7cm}
\end{center}
\caption{\small \label{f13}{Crossing the phantom divide by the
equation of state parameter for $a(t)=a_{0}e^{\nu t}$ and
$\varphi=\varphi_0 e^{-\alpha t}$, where $a_{0}$, $\varphi_0$,
$\nu>0$ and $\alpha>0$ are constants. This crossing of the phantom
divide occurs at $z\approx0.25$ }}
\end{figure}

\Fref{f13} shows the possibility of phantom divide crossing by the
equation of state parameter of the model. The universe transits into
the phantom phase from a quintessence phase in a redshift that is
observationally viable ($z\approx 0.25$).\\

From another perspective, the classical stability of the solutions
in $w_{\varphi}-w'_{\varphi}$ phase-plane gives some interesting
results. To do this end, we calculate $w_{\varphi}$ versus
$w'_{\varphi}$ as

$$
w_{\varphi}^{\prime}= \Bigg[4 \sqrt{6 \xi}(1-6\xi)(1+w_{\varphi})b
x_{1}x_{5}-3(1-6\xi)bw_{\varphi}(1+w_{\varphi})x_{5}^{\,\,2}$$
$$+{\frac{\Big((-6+28\xi)
x_{1}^{\,\,2}+2\sqrt{6}\lambda(1-3\xi)x_{1}x_{2}^{\,\,2}-8\sqrt{6\xi}(1-4\xi)x_{1}x_{5}+2\sqrt{\xi}\lambda
x_{2}^{\,\,2}x_{5}-8\xi
x_{5}^{\,\,2}+2\sqrt{6\xi}\lambda^{2}x_{1}x_{2}^{\,\,2}x_{5}\Big)}{x_{1}^{\,\,2}+x_{2}^{\,\,2}
+x_{5}^{\,\,2}+2\sqrt{6\xi}x_{1}x_{5}}}$$
\begin{equation}
-{\frac{3}{2}}(x_{1}^{\,\,2}+x_{2}^{\,\,2}
+x_{5}^{\,\,2}+2\sqrt{6\xi}x_{1}x_{5})b^{3}(1-6\xi)(1+w_{\varphi})^{2}x_{4}^{\,\,2}x_{5}^{\,\,2}+
3w_{\varphi}(1+w_{\varphi})\Bigg]\Big(1-b(1-6\xi)x_{5}^{\,\,2}\Big)^{-1}
\label{50}
\end{equation}

where by definition
$$b\equiv
\Big(1+\frac{1}{2Hr_{c}}\Big)^{-1}=\Big(1+x_{4}^{2}\Big)^{-1}.$$

\begin{figure}[htp]
\begin{center}\includegraphics{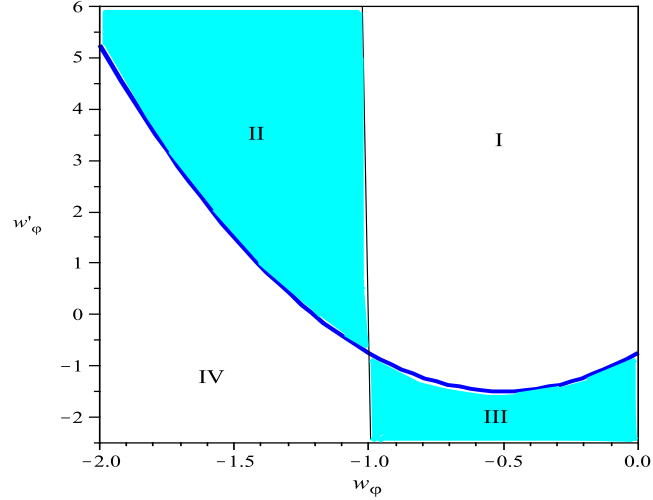} \vspace{6.2cm}
\end{center}
\caption{\small \label{f14}{Bounds on $w_{\varphi}^{\prime}$ as a
function of $w_{\varphi}$ in $w_{\varphi}-w'_{\varphi}$
phase-plane for $\Omega_{\varphi}=0.8$},\, $\xi=1/6$ and
$\lambda=0.1$}
\end{figure}

In this case, the $w_{\varphi}-w'_{\varphi}$ phase-plane is divided
into the following four regions

\begin{equation}
\left\{\begin{array}{ll}w_{\varphi}>-1 \,,\quad w'_{\varphi}>3w_{\varphi}(1+w_{\varphi})&\Longrightarrow c_{a}^{2}<0\,\quad(region\, I)\\ \\
w_{\varphi}<-1 \,,\quad w'_{\varphi}>3w_{\varphi}(1+w_{\varphi})&\Longrightarrow c_{a}^{2}>0\,\quad(region\, II)\\ \\
w_{\varphi}>-1 \,, \quad
w'_{\varphi}<3w_{\varphi}(1+w_{\varphi})&\Longrightarrow{\rm
c_{a}^{2}>0}\,\quad (region\, III)\\ \\
w_{\varphi}<-1 \,,\quad
w'_{\varphi}<3w_{\varphi}(1+w_{\varphi})&\Longrightarrow
c_{a}^{2}<0\,\quad(region\, IV)\label{51}
\end{array}\right.
\end{equation}
As we have explained in previous sections, the stability of the
solutions requires $c_{a}^{2}>0$. So, the stability regions of the
solutions in this case are the regions \emph{II} and \emph{III}. The
region  \emph{II} corresponds to an effective phantom phase while
region \emph{III} is a quintessence phase. \Fref{f14} shows these
regions in $w_{\varphi}-w'_{\varphi}$ phase-plane.

\subsection{Non-minimally coupled phantom field on the normal DGP setup}

For completeness of our analysis, now we consider a non-minimally
coupled phantom field on the normal DGP setup. The energy density
and pressure of this non-minimally coupled phantom field are given
by

\begin{equation}
\rho_{\varphi}=-\frac{1}{2}{\dot{\varphi}^{2}}+V(\varphi)+6\xi
H\varphi{\dot{\varphi}}+3\xi H^{2}{\varphi}^{2}\,,\label{52}
\end{equation}
and
\begin{equation}
p_{\varphi}=-\frac{1}{2}{\dot{\varphi}^{2}}-V(\varphi)-2\xi
(\varphi{\ddot{\varphi}}+2\varphi
H{\dot{\varphi}}+{\dot{\varphi}^{2}})-\xi
{\varphi}^{2}(2\dot{H}+3H^{2})\,.\label{53}
\end{equation}
respectively. As previous cases, the equation governing on the the
evolution of the Hubble parameter depends on the dimensionless
variables and can be written as \\
\begin{equation}
\fl \dot{H}=-{\frac{3(-1-2\xi)x_{1}^{\,\,2}+\sqrt{6\xi}x_{1}
x_{5}+{\frac{3}{2}}\gamma
x_{3}^{\,\,2}+3\sqrt{\xi}x_{5}(\sqrt{6}x_{1}-4\sqrt{\xi}x_{5}+\lambda
x_{2}^{\,\,2})}{1+x_{4}^{\,\,2}-(1+6\xi)x_{5}^{\,\,2}}}H^{2}\,.\label{54}
\end{equation}
The Klien-Gordon equation for a non-minimally coupled phantom field
is
\begin{equation}
\ddot{\varphi}+3H{\dot{\varphi}}-\xi R
\varphi={\frac{dV}{d\varphi}}\,.\label{55}
\end{equation}
The Friedmann equation in terms of the phase space coordinates now
can be written as the following constraint equation
\begin{equation}
-x_{1}^{\,\,2}+x_{2}^{\,\,2}+x_{3}^{\,\,2}+2\sqrt{6\xi}x_{1}x_{5}+x_{5}^{\,\,2}-2x_{4}^{\,\,2}=1\,.\label{56}
\end{equation}
The parameter space of the model as a dynamical system is described
by the following autonomous system
\begin{equation}
\frac{dx_1}{dN}=-3x_1-\frac{\sqrt{6}}{2}\lambda
x_2^{\,\,2}+2\sqrt{6\xi}x_5+(\sqrt{6\xi}x_{5}-x_1)
\Delta\,,\label{57}
\end{equation}
\begin{equation}
\frac{dx_2}{dN}=-\frac{\sqrt{6}}{2}\lambda x_{1} x_{2}-x_{2}
\Delta\,,\label{58}
\end{equation}

\begin{equation}
\frac{dx_4}{dN}=-{\frac{1}{2}}x_{4}\Delta\,,\label{59}
\end{equation}

\begin{equation}
\frac{dx_5}{dN}=\sqrt{6\xi}x_{1}\,\,.\label{60}
\end{equation}
where
$$\Delta\equiv \frac{\dot{H}}{H^{2}}.$$
As previous sections, in order to study the stability of the
critical points, we should obtain their eigenvalues, which are
written as follows:\\

$\bullet$ point (1a),(1b):$$\alpha_{1,2}=\frac{3}{2}\gamma,\quad
\alpha_{3,4}=-\frac{3}{2}+\frac{3}{4} \gamma\pm\frac{1}{4}
\sqrt{36-36 \gamma+9 \gamma^2+192 \xi-144 \xi \gamma}$$

$\bullet$ point (2a),(2b):$$\alpha_{1}=-\frac{3}{2}\gamma+2,\quad
\alpha_{2}=-1,\quad \alpha_{3,4}=2$$

$\bullet$ curve C:$$\alpha_{1}=-\frac{3}{2}\gamma,\quad
\alpha_{2}=0,$$

$$\alpha_{3,4}=\frac{1}{2}\frac{1}{(-16\,\xi-16\,{x_{{2}}}^{2}\xi+{\lambda}^{2}{x_{{2}}}^{4}+12\,{\lambda}^{
2}{x_{{2}}}^{4}\xi)}\Bigg[48\,\xi+48\,{x_{{2}}}^{2}\xi-3\,{\lambda}^{2}{x_{{2}}}^{4}-36\,{
\lambda}^{2}{x_{{2}}}^{4}\xi\,\,$$
$$\pm \Big(2304\,{\xi}^{2}+4608\,{x_{{2}}}^{2}{\xi}^{2}-288\,{\lambda}^{2}{x_{{2}
}}^{4}\xi-1920\,{\lambda}^{2}{x_{{2}}}^{4}{\xi}^{2}+2304\,{x_{{2}}}^{4
}{\xi}^{2}-288\,{x_{{2}}}^{6}\xi\,{\lambda}^{2}$$$$-4992\,{x_{{2}}}^{6}{
\xi}^{2}{\lambda}^{2}+9\,{\lambda}^{4}{x_{{2}}}^{8}+72\,{\lambda}^{4}{
x_{{2}}}^{8}\xi
+1872\,{\lambda}^{4}{x_{{2}}}^{8}{\xi}^{2}+24576\,{x_{{2}}}^{2}{\xi}^{3
}+12288\,{\xi}^{3}$$$$-384\,{\lambda}^{4}{x_{{2}}}^{6}\xi+3072\,{\lambda}^
{2}{x_{{2}}}^{2}{\xi}^{2}+12288\,{x_{{2}}}^{4}{\xi}^{3}+12\,{\lambda}^
{6}{x_{{2}}}^{10}-9216\,{x_{{2}}}^{6}{\xi}^{3}{\lambda}^{2}$$$$-9216\,{
\lambda}^{2}{x_{{2}}}^{4}{\xi}^{3}+144\,{\lambda}^{6}{x_{{2}}}^{10}\xi
-2304\,{\lambda}^{4}{x_{{2}}}^{6}{\xi}^{2}\Big)^\frac{1}{2}\Bigg]$$

\Tref{t5} shows the results of the stability analysis in the phase
space of a non-minimally coupled phantom field. In this case there
are four critical points non of them result in a stable phase for
the system. There is also a critical line, ($C$), which behaves as
line of saddle points.
\begin{table*}
\begin{center}
\caption{\small \label{t5} {Location and dynamical character of
the fixed points.}}
\begin{tabular}{|c|c|c|c|c|c|c|c|c|c|} \hline\mr\ns

 name & $x_{1c}$ & $x_{2c}$&$x_{3c}$ &$x_{5c}$  & stability  &$ \gamma_{\varphi}$&$w_{eff}$&$a(t)$\\
\hline\mr\ns

(1a),(1b) & 0 & 0 & $\pm1$ & 0  & saddle point&undefined &
$\gamma-1$ & $a_{0}(t-t_{0})^{\frac{2}{3\gamma}}$\\
\hline

(2a),(2b) & 0 & 0 & 0 &$ \pm1$  & saddle point  & $4/3$ & $1/3$ &
$a_{0}(t-t_{0})^{\frac{1}{2}}$\\ \hline

C & 0 & $x_{2}$ & 0 & $\frac{\lambda x_{2}^{\,\,2}}{4 \sqrt{\xi}}$
& saddle point & $0$ & $-1$ & $e^{\Lambda (t-t_{0})}$  \\
\hline\mr\ns
\end{tabular}
\end{center}
\end{table*}

\begin{figure}[htp]
\begin{center}\includegraphics{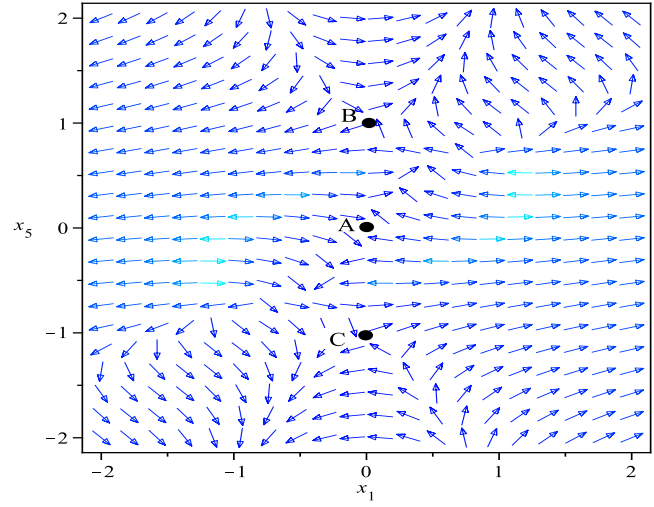} \vspace{7cm}
\end{center}
\caption{\small \label{f017}{ The $x_{1}-x_{5}$ phase plane for a
non-minimally coupled phantom field with $\xi=1/6$. Points $A$, $B$
and $C$ are saddle points. }}
\end{figure}

\Fref{f017} shows the $x_{1}-x_{5}$ phase plane of the model with
$\xi=1/6$. In this case points $A$, $B$ and $C$ behave as saddle
points.

\begin{figure}[htp]
\begin{center}\includegraphics{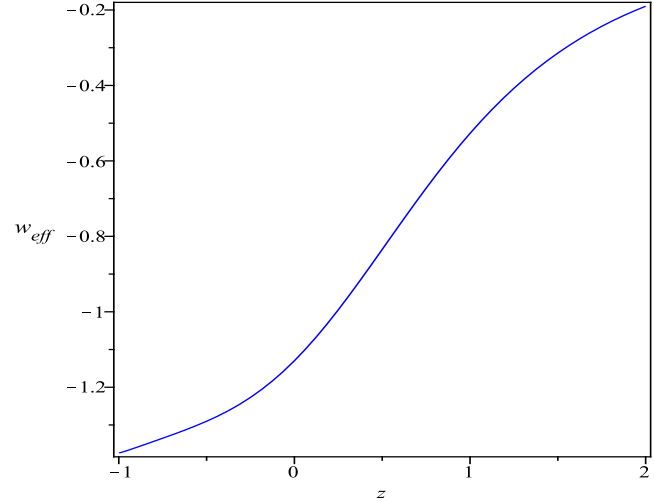} \vspace{7cm}
\end{center}
\caption{\small \label{f15}{ Crossing of the phantom line by the
equation of state parameter for a phantom field on the normal DGP
setup. We considered $a(t)=a_{0}e^{\nu t}$ and $\varphi=\varphi_0
e^{-\alpha t}$, where $a_{0}$, $\varphi_0$, $\nu$ and $\alpha$ are
positive constants. Crossing of the phantom divide occurs at
$z\approx0.24$ }}
\end{figure}

Now let us to study the possibility of phantom divide crossing by
the equation of state parameter in this setup. As \Fref{f15} shows,
in this case transition to the phantom phase occurs from
quintessence phase at $z \approx 0.25$. Finally, investigation of
classical stability of the solutions in $w_{\varphi}-w'_{\varphi}$
phase-plane gives interesting result. We calculate $w_{\varphi}$
versus $w'_{\varphi}$ for this model as
\begin{equation}
w_{\varphi}^{\prime}= \Bigg[4 \sqrt{6 \xi}(1+6\xi)(1+w_{\varphi})b
x_{1}x_{5}-3(1+6\xi)bw_{\varphi}(1+w_{\varphi})x_{5}^{\,\,2}
\end{equation}
$$+{\frac{\Big((6+28\xi)
x_{1}^{\,\,2}+2\sqrt{6}\lambda(1+3\xi)x_{1}x_{2}^{\,\,2}-8\sqrt{6\xi}(1+4\xi)x_{1}x_{5}-2\sqrt{\xi}\lambda
x_{2}^{\,\,2}x_{5}+8\xi
x_{5}^{\,\,2}-2\sqrt{6\xi}\lambda^{2}x_{1}x_{2}^{\,\,2}x_{5}\Big)}{-x_{1}^{\,\,2}+x_{2}^{\,\,2}
+x_{5}^{\,\,2}+2\sqrt{6\xi}x_{1}x_{5}}}$$
$$-{\frac{3}{2}}(-x_{1}^{\,\,2}+x_{2}^{\,\,2}
+x_{5}^{\,\,2}+2\sqrt{6\xi}x_{1}x_{5})b^{3}(1+6\xi)(1+w_{\varphi})^{2}x_{4}^{\,\,2}x_{5}^{\,\,2}+
3w_{\varphi}(1+w_{\varphi})\Bigg]\Big(1-b(1+6\xi)x_{5}^{\,\,2}\Big)^{-1}$$\\\label{61}

\begin{figure}[htp]
\begin{center}\includegraphics{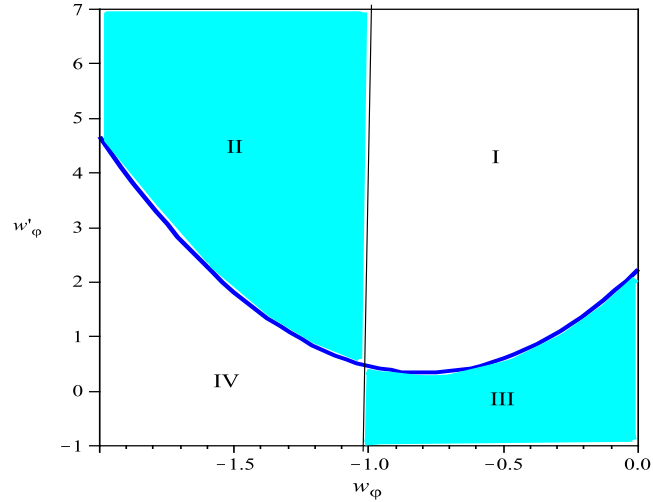} \vspace{7cm}
\end{center}
\caption{\small \label{f16}{Bounds on $w_{\varphi}^{\prime}$ as a
function of $w_{\varphi}$ in $w_{\varphi}-w'_{\varphi}$ phase-plane
for $\Omega_{\varphi}=0.8$}, $\xi=1/6$ and $\lambda=1$. The
stability regions of the phase-plane are painted with cyan color.}
\end{figure}

The $w_{\varphi}-w'_{\varphi}$ phase-plane of the model is plotted
in \fref{f16}. This phase-plane can be divided into the following
four regions

\begin{equation}
\left\{\begin{array}{ll}w_{\varphi}>-1 \,,\quad w'_{\varphi}>3w_{\varphi}(1+w_{\varphi})&\Longrightarrow c_{a}^{2}<0\,\quad(region\, I)\\ \\
w_{\varphi}<-1 \,,\quad w'_{\varphi}>3w_{\varphi}(1+w_{\varphi})&\Longrightarrow c_{a}^{2}>0\,\quad(region\, II)\\ \\
w_{\varphi}>-1 \,, \quad
w'_{\varphi}<3w_{\varphi}(1+w_{\varphi})&\Longrightarrow{\rm
c_{a}^{2}>0}\,\quad (region\, III)\\ \\
w_{\varphi}<-1 \,,\quad
w'_{\varphi}<3w_{\varphi}(1+w_{\varphi})&\Longrightarrow
c_{a}^{2}<0\,\quad(region\, IV)
\end{array}\right.\label{62}
\end{equation}
The stable regions of the phase-plane are those regions that the
condition $c_{a}^{2}>0$ is fulfilled. In this case, these stable
regions are Region II and III as shown in \fref{f16}.

\section{Summary and Conclusion}
The accelerated expansion of the universe supported by recent
observational data could be associated with dark energy, whose
theoretical nature and origin are still unknown for cosmologists.
Cosmological constant or vacuum energy with an equation of state
parameter $\omega=-1$, is the most popular candidate for dark
energy, but unfortunately it suffers from some serious problems such
as huge fine-tuning and coincidence problems. Therefore, a number of
models containing dynamical dark energy have been proposed as
responsible mechanisms for late-time cosmic speed up. Some of these
models are quintessence, k-essence, phantom scalar field, chaplygin
gas models and so on. Another alternative approach to explain the
late-time cosmic speed up is modification of the geometric sector of
the Einstein field equations leading to modified gravity theories.
In the spirit of modified gravity proposal, the
Dvali-Gabadadze-Porrati (DGP) braneworld scenario explains the
late-time accelerated expansion in its self-accelerating branch
without need to introduce a dark energy component on the brane.
However, some important features of dark energy such as possible
crossing of the cosmological constant equation of state parameter
are missing in the pure DGP model. In addition, the
self-accelerating DGP solution suffers from ghost instability which
makes the model unfavorable. Incorporation of a scalar field
component on the DGP brane and treating the normal branch solutions
brings a lots of new physics, some of which are studied in this
paper. Previous studies in this field are restricted to either
scalar fields on the self-accelerating branch or the simple case of
minimally coupled scalar fields. In this paper we considered a
scalar field component (both quintessence and phantom scalar
fields), non-minimally coupled with induced gravity on the brane. We
studied cosmological dynamics of the normal branch solutions on the
brane within a dynamical system approach. We translated dynamical
equations into an autonomous dynamical system in each case. Then we
obtained the critical points of the model in phase space of each
model. The issue of stability of these solutions are studied with
details. Also possibility of having a stable attractor in de Sitter
phase corresponding to current accelerated phase of universe
expansion are studied in each case. We have also investigated the
possibility to have a transition to the phantom phase of the
equation of state parameter in each case. The classical stability of
the solutions are treated also in a $w_{\varphi}-w'_{\varphi}$
phase-plane analysis in each step. We have shown in each step that
there are several phases of accelerated expansion in each case, but
only a limited critical points have stable, attractor solutions with
de Sitter scale factor describing the current accelerated expansion
on the brane. In summary, the main achievements of this study are as
follows: \\

$\bullet$ While the pure, normal DGP solution has not the potential
to explain late-time cosmic acceleration, with a minimally coupled
quintessence field in the normal DGP setup there is a stable de
Sitter phase realizing the late-time cosmic speed up. Nevertheless,
as the pure DGP case, there is no possibility to cross the
cosmological constant line by the effective equation of state
parameter of the model. The classical stability domain of the model
is restricted to those subspaces of the model parameter space that
$w'_{\varphi}<3w_{\varphi}(1+w_{\varphi})$  with $w_{\varphi}>-1$
where
$w'_{\varphi}\equiv\frac{dw_{\varphi}}{dN}$ and $N\equiv\ln a(t)$.\\

$\bullet$ With a minimally coupled phantom field on the brane, it is
possible to have an attractor, de Sitter solution realizing the
late-time accelerated expansion in the normal DGP setup. Also, the
effective equation of state parameter of the model crosses the
phantom divide. Note that this crossing is impossible by the
effective equation of state parameter of a minimally coupled
quintessence field on the brane. Similar to the minimally
quintessence field on the brane, the classical stability domain of
the model is restricted to those subspaces of the model parameter
space that $w'_{\varphi}<3w_{\varphi}(1+w_{\varphi})$ with
$w_{\varphi}<-1$.\\

$\bullet$ With a non-minimally coupled quintessence scalar field on
the DGP brane, it is possible to realize a stable, de Sitter
late-time accelerating phase in the normal branch of the model. It
is possible also to cross the phantom divide by the effective
equation of state parameter of the model. In this case, there are
two different domains of stability in the $w_{\varphi}-w'_{\varphi}$
phase-plane of the model: a subspace with
$w'_{\varphi}>3w_{\varphi}(1+w_{\varphi})$ with $w_{\varphi}<-1$
corresponding to an \emph{effective phantom phase} and the other
subspace with $w'_{\varphi}<3w_{\varphi}(1+w_{\varphi})$ with
$w_{\varphi}>-1$ corresponding to a quintessence phase. This feature
shows that it is possible to have an effective phantom picture with
a non-minimally quintessence field on the normal DGP setup. \\

$\bullet$ For a non-minimally coupled phantom scalar field on the
DGP brane, it is possible to realize a de Sitter late-time
accelerating phase in the normal branch of the model. It is possible
also to cross the phantom divide line by the effective equation of
state parameter of the model in this case. Also, as for the case of
non-minimally coupled quintessence field on the brane, there are two
different domains of stability in the $w_{\varphi}-w'_{\varphi}$
phase-plane of the model: a subspace with
$w'_{\varphi}>3w_{\varphi}(1+w_{\varphi})$ with $w_{\varphi}<-1$
corresponding to the phantom phase and the other subspace with
$w'_{\varphi}<3w_{\varphi}(1+w_{\varphi})$ with $w_{\varphi}>-1$
corresponding to an \emph{effective quintessence phase} on the
brane. Note that it is possible to have an effective quintessence
picture with a non-minimally coupled phantom field on the normal DGP
setup.

Finally we stress that the observational status of the present
DGP-inspired models can be treated in the same line as has been
reported in Ref. \cite{25}.

\end{document}